\PassOptionsToPackage{unicode}{hyperref}
\PassOptionsToPackage{hyphens}{url}
\PassOptionsToPackage{dvipsnames,svgnames,x11names}{xcolor}
\documentclass[
  american,
  12pt]{article}
\usepackage{xcolor}
\usepackage{amsmath,amssymb}
\usepackage{iftex}
\ifPDFTeX
  \usepackage[T1]{fontenc}
  \usepackage[utf8]{inputenc}
  \usepackage{textcomp} % provide euro and other symbols
\else % if luatex or xetex
  \usepackage{unicode-math} % this also loads fontspec
  \defaultfontfeatures{Scale=MatchLowercase}
  \defaultfontfeatures[\rmfamily]{Ligatures=TeX,Scale=1}
\fi
\usepackage{lmodern}
\ifPDFTeX\else
\fi
\IfFileExists{upquote.sty}{\usepackage{upquote}}{}
\IfFileExists{microtype.sty}{% use microtype if available
  \usepackage[]{microtype}
  \UseMicrotypeSet[protrusion]{basicmath} % disable protrusion for tt fonts
}{}
\makeatletter
\@ifundefined{KOMAClassName}{% if non-KOMA class
  \IfFileExists{parskip.sty}{%
    \usepackage{parskip}
  }{% else
    \setlength{\parindent}{0pt}
    \setlength{\parskip}{6pt plus 2pt minus 1pt}}
}{% if KOMA class
  \KOMAoptions{parskip=half}}
\makeatother
\makeatletter
\ifx\paragraph\undefined\else
  \let\oldparagraph\paragraph
  \renewcommand{\paragraph}{
    \@ifstar
      \xxxParagraphStar
      \xxxParagraphNoStar
  }
  \newcommand{\xxxParagraphStar}[1]{\oldparagraph*{#1}\mbox{}}
  \newcommand{\xxxParagraphNoStar}[1]{\oldparagraph{#1}\mbox{}}
\fi
\ifx\subparagraph\undefined\else
  \let\oldsubparagraph\subparagraph
  \renewcommand{\subparagraph}{
    \@ifstar
      \xxxSubParagraphStar
      \xxxSubParagraphNoStar
  }
  \newcommand{\xxxSubParagraphStar}[1]{\oldsubparagraph*{#1}\mbox{}}
  \newcommand{\xxxSubParagraphNoStar}[1]{\oldsubparagraph{#1}\mbox{}}
\fi
\makeatother

\usepackage{longtable,booktabs,array}
\usepackage{calc} % for calculating minipage widths
\usepackage{etoolbox}
\makeatletter
\patchcmd\longtable{\par}{\if@noskipsec\mbox{}\fi\par}{}{}
\makeatother
\IfFileExists{footnotehyper.sty}{\usepackage{footnotehyper}}{\usepackage{footnote}}
\makesavenoteenv{longtable}
\usepackage{graphicx}
\makeatletter
\newsavebox\pandoc@box
\newcommand*\pandocbounded[1]{% scales image to fit in text height/width
  \sbox\pandoc@box{#1}%
  \Gscale@div\@tempa{\textheight}{\dimexpr\ht\pandoc@box+\dp\pandoc@box\relax}%
  \Gscale@div\@tempb{\linewidth}{\wd\pandoc@box}%
  \ifdim\@tempb\p@<\@tempa\p@\let\@tempa\@tempb\fi% select the smaller of both
  \ifdim\@tempa\p@<\p@\scalebox{\@tempa}{\usebox\pandoc@box}%
  \else\usebox{\pandoc@box}%
  \fi%
}
\def\fps@figure{htbp}
\makeatother

\ifLuaTeX
\usepackage[bidi=basic]{babel}
\else
\usepackage[bidi=default]{babel}
\fi
\def\languageshorthands#1{}
\ifLuaTeX
  \usepackage[english]{selnolig} % disable illegal ligatures
\fi

\providecommand{\tightlist}{%
  \setlength{\itemsep}{0pt}\setlength{\parskip}{0pt}}

\usepackage[]{natbib}
\usepackage{tikz}
\usetikzlibrary{decorations.pathreplacing}
\usetikzlibrary{positioning}
\usetikzlibrary{shapes.arrows,shapes.symbols,arrows.meta}
\usepackage[no-weekday]{eukdate}
\usepackage{bm}
\usepackage{mathptmx}
\usepackage{amsmath} % Ensure amsmath is included for \eqref
\usepackage{bbm}
\usepackage[]{natbib}
\allowdisplaybreaks
\expandafter\def\expandafter\normalsize\expandafter{%
  \normalsize
  \setlength\abovedisplayskip{6pt}
  \setlength\belowdisplayskip{6pt}
  \setlength\abovedisplayshortskip{3pt}
  \setlength\belowdisplayshortskip{3pt}
}
\usepackage{microtype}
\usepackage[onlyrefs]{mathtools}
\usepackage{threeparttable}
\makeatletter
\g@addto@macro\TPT@defaults{\linespread{1}\selectfont}
\def\fps@table{!tbh}
\def\fps@figure{!tbh}
\makeatother

\usepackage{setspace}
\usepackage{caption}
\DeclareCaptionStyle{italic}[justification=centering]
 {labelfont={bf},textfont={it},labelsep=colon}
\usepackage{titlesec}
\titlespacing*{\section}
  {0pt}{0.3\baselineskip}{0.1\baselineskip}
\titlespacing*{\subsection}
  {0pt}{0.3\baselineskip}{0.1\baselineskip}
\makeatletter
\@ifpackageloaded{caption}{}{\usepackage{caption}}
\AtBeginDocument{%
\ifdefined\contentsname
  \renewcommand*\contentsname{Table of contents}
\else
  \newcommand\contentsname{Table of contents}
\fi
\ifdefined\listfigurename
  \renewcommand*\listfigurename{List of Figures}
\else
  \newcommand\listfigurename{List of Figures}
\fi
\ifdefined\listtablename
  \renewcommand*\listtablename{List of Tables}
\else
  \newcommand\listtablename{List of Tables}
\fi
\ifdefined\figurename
  \renewcommand*\figurename{Figure}
\else
  \newcommand\figurename{Figure}
\fi
\ifdefined\tablename
  \renewcommand*\tablename{Table}
\else
  \newcommand\tablename{Table}
\fi
}
\@ifpackageloaded{float}{}{\usepackage{float}}
\floatstyle{ruled}
\@ifundefined{c@chapter}{\newfloat{codelisting}{h}{lop}}{\newfloat{codelisting}{h}{lop}[chapter]}
\floatname{codelisting}{Listing}

\usepackage{amsthm}
\theoremstyle{plain}
\newtheorem{corollary}{Corollary}[section]
\theoremstyle{plain}
\newtheorem{proposition}{Proposition}[section]
\theoremstyle{remark}
\AtBeginDocument{}

\makeatother
\makeatletter
\@ifpackageloaded{caption}{}{\usepackage{caption}}
\@ifpackageloaded{subcaption}{}{\usepackage{subcaption}}
\makeatother
\usepackage{bookmark}
\IfFileExists{xurl.sty}{\usepackage{xurl}}{} % add URL line breaks if available
\hypersetup{
  pdftitle={Online conformal inference for multi-step time series forecasting},
  pdfauthor={Xiaoqian Wang; Rob J Hyndman},
  pdflang={en-US},
  pdfkeywords={Conformal prediction, Coverage
guarantee, Distribution-free inference, Exchangeability, Delayed
feedback},
  colorlinks=true,
  linkcolor={blue},
  filecolor={Maroon},
  citecolor={Blue},
  urlcolor={Blue},
  pdfcreator={LaTeX via pandoc}}

\begin{document}
\def\spacingset#1{\renewcommand{\baselinestretch}%
{#1}\small\normalsize} \spacingset{1}

\renewcommand*{\arraystretch}{0.5} % Specify row height in a table globally

%%%%%%%%%%%%%%%%%%%%%%%%%%%%%%%%%%%%%%%%%%%%%%%%%%%%%%%%%%%%%%%%%%%%%%%%%%%%%%

\title{\bf Online conformal inference for multi-step time series
forecasting}
\author{
Xiaoqian Wang\thanks{Corresponding author. Xiaoqian Wang, Academy of
Mathematics and Systems Science, Chinese Academy of Sciences, Beijing,
100190, China. E-mail address:
\href{mailto:xiaoqian.wang@amss.ac.cn}{xiaoqian.wang@amss.ac.cn} (X.
Wang).} \vspace{0.2em}\\
Academy of Mathematics and Systems Science, Chinese Academy of
Sciences \vspace{0.2em}\\
and \vspace{0.2em}\\Rob J Hyndman \vspace{0.2em}\\
Department of Econometrics \& Business Statistics, Monash
University \vspace{0.2em}\\
}
\maketitle

\bigskip
\bigskip
\begin{abstract}
We study the construction of distribution-free prediction intervals for
multi-step time series forecasting. In this setting, forecast errors at
different horizons are serially dependent, and feedback is delayed,
since the error of an \(h\)-step-ahead forecast becomes available only
after \(h\) periods. We propose autocorrelated multi-step conformal
prediction (AcMCP), an online procedure that accommodates delayed
feedback and explicitly incorporates the dependence among multi-step
forecast errors into conformal calibration. To characterize this
dependence, we derive first-order representations of the optimal
forecast errors under a general nonstationary autoregressive
data-generating process, with explicit bounds on remainder terms. AcMCP
guarantees long-run coverage of at least the nominal level at every
horizon, without requiring exchangeability or any assumption on how the
data distribution evolves over time. We also establish a finite-sample
bound on the empirical coverage shortfall, which depends on the forecast
horizon and the evaluation sample size. Notably, these guarantees do not
rely on the accuracy of the linear approximations. Simulations and
real-data applications show that AcMCP adapts to time-varying dynamics
and maintains rolling coverage close to the nominal level. By exploiting
the dependence among multi-step errors, it often yields narrower
intervals than its counterpart without this dependence, with similar
coverage, particularly at longer horizons.
\end{abstract}

\noindent%
\vfill

\newpage
\spacingset{1.8} % DON'T change the spacing!

\renewcommand{\theproposition}{\arabic{proposition}} % numbering propositions sequentially and individually by type and not by section
\setcounter{proposition}{0}

\renewcommand{\thecorollary}{\arabic{corollary}} % numbering corollaries sequentially and individually
\counterwithout{corollary}{section}
\setcounter{corollary}{0}

\section{Introduction}\label{sec-intro}

Conformal prediction \citep{papadopoulos2002, vovk2005} is a
model-agnostic framework for uncertainty quantification that constructs
prediction intervals with valid coverage, without requiring stringent
assumptions on the underlying data distribution. Its classical
guarantees, however, rely on the data being i.i.d. or, more generally,
exchangeable. In time series forecasting, observations are typically
serially dependent and may exhibit nonstationarity or distributional
shifts, so exchangeability generally fails and the standard guarantees
no longer hold.

Recent work has sought to relax this requirement. One line of research
replaces exchangeability with weaker, time-series-specific assumptions.
To handle temporal dependence, approximate coverage guarantees have been
obtained through permutation schemes that preserve the temporal
dependence structure \citep{chernozhukov2018, yu2022} and through
concentration and decoupling arguments for split conformal prediction
\citep{oliveira2024}. \citet{xu2021} use bootstrap ensemble predictors
to obtain out-of-bag residuals and update their empirical distribution
as new observations arrive, establishing approximate coverage under
stationary, strongly mixing errors. Under covariate shift,
\citet{tibshirani2019} and \citet{yang2024} establish coverage
guarantees when the training and test distributions differ only in their
covariate marginals. Other work generalizes conformal calibration by
replacing the unweighted empirical distribution of nonconformity scores
with a weighted one \citep{barber2023, guan2023}.

A separate strand of research develops online conformal methods that
adjust prediction intervals using sequential coverage feedback. Adaptive
conformal inference \citep[ACI,][]{gibbs2021} updates the working
miscoverage level in response to observed coverage errors, so that
empirical coverage converges to the target level without distributional
assumptions on the time series. Subsequent extensions use expert
aggregation and adaptive online learning to reduce sensitivity to the
learning rate and to improve adaptation to distribution shifts
\citep{zaffran2022, gibbs2024, bhatnagar2023}. Conformal PID control
\citep{angelopoulos2024} builds on the same feedback principle but
adjusts the thresholds of the nonconformity scores directly, combining
quantile tracking, correction for cumulative coverage errors, and a
scorecaster that exploits predictable temporal structure in the scores.
Further refinements improve stability through decaying step sizes
\citep{angelopoulos2024online}, reduce tuning through parameter-free
updates \citep{podkopaev2024}, and incorporate smooth feedback that
reflects the magnitude of the discrepancy between scores and thresholds
\citep{wu2025}.

Most of this literature, however, targets single-step forecasting,
whereas many applications require uncertainty quantification over
several future horizons. Multi-step forecasting poses two additional
challenges for conformal methods. First, feedback is delayed: the
realized error of an \(h\)-step-ahead forecast becomes available only
\(h\) periods after the forecast is issued, so online updates must rely
on outdated information. Second, forecast errors at different horizons
issued from the same origin are dependent, because errors propagate
through the forecasting recursion.

Existing conformal methods for multi-step forecasting address these
challenges only partially. \citet{stankeviciute2021} integrate conformal
prediction with recurrent neural networks for multi-step forecasting and
apply a Bonferroni correction to obtain simultaneous coverage across
horizons, but their guarantee relies on exchangeability across time
series. \citet{schlembach2025} extend this approach to multivariate
targets under nonexchangeability, with a joint miscoverage bound across
horizons and response dimensions. Bellman conformal inference
\citep{yang2024ts} uses multi-step forecasts in a look-ahead
optimization that trades off interval length against a miscoverage
penalty, but it outputs, and guarantees coverage for, only the resulting
sequence of one-step-ahead intervals; it also requires nested interval
families, which calls for additional construction when only point
forecasts are available, and solves a dynamic program at every time
step. The multi-step extension of sequential predictive conformal
inference \citep[SPCI,][]{xu2023} fits a separate predictor and
conditional quantile model for each horizon and uses past residuals to
adapt the interval bounds; although SPCI has asymptotic conditional
coverage guarantees under suitable assumptions, its multi-step
implementation is not analyzed separately. To our knowledge, existing
multi-step conformal methods seldom provide coverage guarantees without
exchangeability, and none exploits the dependence among forecast errors
across horizons. A method that handles delayed feedback, accounts for
this dependence, and retains distribution-free coverage guarantees is
therefore still lacking.

We consider a time series \(\{y_t\}_{t\ge1}\) generated by an unknown,
possibly nonstationary data-generating process (DGP) that may depend on
its own past and on exogenous predictors \(\bm{x}_t\in\mathbb{R}^p\). At
each time \(t\), we wish to construct a prediction interval
\(\hat{\mathcal{C}}_{t+h|t}\) for \(y_{t+h}\) at every horizon
\(h\in[H]\). Given a user-specified miscoverage level
\(\alpha\in(0,1)\), the goal is horizon-specific long-run coverage,
namely
\(\liminf_{T\to\infty}\frac{1}{T}\sum_{t=1}^{T}\mathbbm{1}\left\{y_{t+h}\in\hat{\mathcal{C}}_{t+h|t}\right\}\geq 1-\alpha\),
for \(h\in[H]\).

This paper makes three contributions. First, we develop multi-step
conformal prediction methods within an online learning framework with
sequential training and calibration splits. Building on conformal PID
control of \citet{angelopoulos2024}, we introduce its multi-step
extension (MPID), which accommodates delayed coverage feedback. Second,
we propose autocorrelated multi-step conformal prediction (AcMCP), which
explicitly incorporates the dependence among multi-step forecast errors
into the scorecaster. To motivate its design, we analyze optimal
forecast errors under a general nonstationary autoregressive DGP with
future exogenous predictors treated as known, derive first-order error
decompositions, and bound the Taylor remainders under suitable
smoothness and moment conditions. These results lead to a scorecaster
that combines a moving average model for errors at a fixed horizon with
a regression on errors across horizons from the same forecast origin.
AcMCP combines these error forecasts with online calibration of the two
tails to construct asymmetric prediction intervals for each horizon.
Third, for both MPID and AcMCP, we establish finite-sample bounds on
empirical coverage that explicitly account for delayed feedback and
imply long-run coverage of at least the nominal level at each fixed
horizon. These guarantees hold under suitable conditions on the scores
and the calibration procedure, without any assumption on how the data
distribution changes over time. They concern coverage averaged over time
at each horizon, rather than simultaneous coverage across horizons or
conditional coverage at individual forecast origins. Importantly, the
coverage guarantees of AcMCP do not depend on the accuracy of the
forecast-error approximations; the contribution of the scorecaster to
interval efficiency is assessed empirically. Simulations and
applications to electricity demand and eating-out expenditure show that
AcMCP adapts to distributional changes and often produces narrower
intervals than MPID with similar coverage, particularly at longer
horizons. A post-hoc comparison shows that this advantage does not
extend uniformly to an adaptive benchmark whose step size is tuned on
the test set.

We developed the \texttt{conformalForecast} package for R to implement
the proposed multi-step conformal prediction methods. The package is
publicly available on CRAN. All the data and code to reproduce the
experiments are made available at \url{https://github.com/xqnwang/cpts}.

\section{Online learning with sequential splits}\label{sec-setup}

\textbf{Problem setup.} We observe a time series \(\{y_t\}_{t\geq1}\)
generated by an unknown DGP, which may depend on its own past as well as
on exogenous predictors
\(\bm{x}_t=(x_{1,t},\dots,x_{p,t})^{\prime}\in\mathbb{R}^p\) and their
histories. Write \(z_t=(\bm{x}_t,y_t)\). The joint distribution of
\(\{z_t\}_{t\geq1}\) is allowed to vary over time, thereby accommodating
nonstationary processes, and we impose no assumption on how it evolves.
At each time \(t\), we forecast \(H\) steps ahead and, for each
\(h\in[H]\), construct a \emph{prediction set}
\(\hat{\mathcal{C}}^{\alpha}_{t+h|t}\) for \(y_{t+h}\), which is a
prediction interval in this setting. The \(h\)-step-ahead forecast uses
the observed data \(\{z_i\}_{1\leq i\leq t}\) together with the
exogenous predictors \(\{\bm{x}_{t+j}\}_{1\leq j\leq h}\) over the
forecast period. These future predictors may be replaced by forecasts
based on information available up to and including time \(t\)
(\emph{ex-ante} forecasts), or the actual values may be assumed
available (\emph{ex-post} forecasts). Given a user-specified
\emph{miscoverage level} \(\alpha\in(0,1)\), we seek intervals with
horizon-specific long-run coverage,
\begin{equation}\phantomsection\label{eq-longrun}{
\liminf_{T\to\infty}\frac{1}{T-h}\sum_{t=h+1}^{T}\mathbbm{1}\left\{y_t\in\hat{\mathcal{C}}^{\alpha}_{t|t-h}\right\}\geq 1-\alpha,\qquad h\in[H].
}\end{equation} This criterion concerns coverage averaged over time at
each forecast horizon, rather than coverage conditional on the available
information or simultaneous coverage of the entire future path. The
parametric form of the DGP is assumed only for the forecast-error
analysis in Section~\ref{sec-ppt}; the coverage guarantees in
Section~\ref{sec-cov} do not require it.

\textbf{Forecasts and errors.} Suppose that, at each time \(t\), a
forecasting model \(\hat{f}_t\) is trained on historical data
\(z_{1:t}\) and updated (retrained) as new observations arrive. Using
\(\hat{f}_t\), we produce point forecasts \(\hat{y}_{t+h|t}\) and signed
forecast errors \(e_{t+h|t}=y_{t+h}-\hat{y}_{t+h|t}\) for \(h\in[H]\).
Conformal inference then calibrates these errors to turn the point
forecasts into prediction intervals.

\textbf{Prediction intervals.} We construct generally asymmetric
intervals by calibrating the lower and upper tails separately. Choose
tail levels \(\alpha^{-},\alpha^{+}>0\) satisfying
\(\alpha^{-}+\alpha^{+}=\alpha\). The two calibration procedures use the
\emph{nonconformity scores} \(s_{t+h|t}^{-}=-e_{t+h|t}\) and
\(s_{t+h|t}^{+}=e_{t+h|t}\), respectively, and produce thresholds
\(q_{t+h|t}^{-}\) and \(q_{t+h|t}^{+}\). These thresholds define the
prediction interval
\begin{equation}\phantomsection\label{eq-two-sided-pi}{
\hat{\mathcal{C}}_{t+h|t}^{\alpha}
:=\left[\hat y_{t+h|t}-q_{t+h|t}^{-},\ \hat y_{t+h|t}+q_{t+h|t}^{+}\right].
}\end{equation} For each tail \(a\in\{-,+\}\), the \emph{miscoverage
indicator}
\(\mathrm{err}_{t+h|t}^{a}:=\mathbbm{1}\{s_{t+h|t}^{a}>q_{t+h|t}^{a}\}\)
records whether the observation falls beyond the corresponding interval
bound. Using opposite signed scores allows both tails to be calibrated
through the same score-exceedance rule, with observations at either
endpoint counted as covered. Because \(y_{t+h}\), and hence
\(s_{t+h|t}^{a}\) and \(\mathrm{err}_{t+h|t}^{a}\), is revealed only at
time \(t+h\), feedback is delayed: at time \(t\), the information
available for horizon \(h\) comes from forecast origins \(t-h\) and
earlier. We use \(\alpha^{-}=\alpha^{+}=\alpha/2\) throughout the
empirical analysis.

Unless otherwise stated, we present the methods and theoretical results
in subsequent sections for a generic single tail, omitting the tail
superscript for simplicity. Accordingly, \(\alpha\) denotes the target
tail miscoverage level in single-tail statements and the total
miscoverage level in two-sided statements. For a candidate response
\(y\), \(s_{t+h|t}^{y}\) denotes the score obtained by replacing
\(y_{t+h}\) with \(y\).

\textbf{Sequential split.} In a time series context, it is inappropriate
to perform \emph{random splitting}, a standard strategy in much of the
conformal prediction literature, due to the temporal dependency present
in the data. Following related work such as \citet{wisniewski2020} and
\citet{zaffran2022}, all conformal methods proposed in this paper use a
\emph{sequential split} to preserve the temporal structure. For example,
the \(t\) available data points, \(z_{1:t}\), are sequentially split
into two consecutive sets, a \emph{proper training set}
\(\mathcal{D}_{\text{tr}} \subset \{1, \dots, t_r\}\) and a
\emph{calibration set}
\(\mathcal{D}_{\text{cal}} \subset \{t_r+1, \dots, t\}\), where
\(t_c=t-t_r \gg H\),
\(\mathcal{D}_{\text{tr}} \cup \mathcal{D}_{\text{cal}} = \{1, \dots, t\}\),
and
\(\mathcal{D}_{\text{tr}} \cap \mathcal{D}_{\text{cal}} = \varnothing\).
Here, ``proper'' means that the training set is used exclusively for
fitting the model, with no overlap with the calibration set, which is
essential for the validity of coverage in conformal prediction
\citep{papadopoulos2002, vovk2005}. With sequential splitting, multiple
\(H\)-step forecasts and their nonconformity scores can be computed on
the calibration set.

\textbf{Online learning.} We adopt the following generic online learning
framework, closely related to \citet{zaffran2022}, for all conformal
methods discussed in later sections. The procedure is illustrated in
\autoref{fig-flowchart}. This framework updates prediction intervals as
new data points arrive, allowing us to assess their long-run coverage
behavior. It adopts a standard rolling window evaluation strategy,
although expanding windows could easily be used instead.

\begin{figure}[!hbtp]
\vspace*{-1.5cm}
\hspace*{-2.5cm}
\centering
\scalebox{0.85}{
  \input figs/annotation.tex
}
\vspace*{-0.5cm}

\caption{Online learning framework with sequential splits. White: unused data; Gray: training data; Black: forecasts in the calibration set; Blue: forecasts in the test set.}\label{fig-flowchart}
\end{figure}

\begin{enumerate}
\def\labelenumi{\arabic{enumi}.}
\item
  \emph{Initialization}. Train a forecasting model on the initial proper
  training set \(z_{(1+t-t_r):t}\), setting \(t=t_r\). Then generate
  \(H\)-step-ahead point forecasts \(\{\hat{y}_{t+h|t}\}_{h\in[H]}\) and
  compute the corresponding upper- and lower-tail nonconformity scores
  \(\{s_{t+h|t}^{+},s_{t+h|t}^{-}\}_{h\in[H]}\) based on the true values
  \(\{y_{t+h}\}_{h\in[H]}\).
\item
  \emph{Recurring procedure}. Roll the training set forward by one data
  point by setting \(t \rightarrow t+1\). Then repeat step 1 until the
  nonconformity scores on the entire initial calibration set,
  \(\{s_{t+h|t}^{+},s_{t+h|t}^{-}\}_{t_r \leq t \leq t_r+t_c-h}\) for
  \(h\in[H]\), are computed.
\item
  \emph{Quantile estimation and prediction interval calculation}. For
  each tail, use the nonconformity scores from the calibration set to
  estimate horizon-specific quantiles. Combine the two one-sided
  thresholds via Equation \eqref{eq-two-sided-pi} to compute
  \(H\)-step-ahead prediction intervals on the test set.
\item
  \emph{Online updating}. Continuously roll the training and calibration
  sets forward, one data point at a time, to update the nonconformity
  scores used for calibration, and repeat step 3 until prediction
  intervals for the entire test set are obtained. These intervals are
  \(\{\hat{\mathcal{C}}_{t+h|t}^{\alpha}\}_{t_r+t_c \leq t \leq T-H}\)
  for \(h \in [H]\), where \(T-t_r-t_c\) is the length of the test set
  used to evaluate coverage against the long-run target in Equation
  \eqref{eq-longrun}.
\end{enumerate}

\section{Multi-step conformal PID control}\label{sec-mpid}

For a well-specified forecasting model, the conditional mean minimizes
the expected squared forecast error, and for covariance-stationary
linear processes the variance of the resulting error \(e_{t+h|t}\) is
non-decreasing in \(h\) \citep{diebold1996, patton2007}. The
distribution of nonconformity scores therefore differs across horizons,
so the thresholds should be calibrated separately for each \(h\in[H]\).
Calibration also need to cope with delayed feedback: at time \(t\), the
realized \(h\)-step score is available only for forecasts issued at
origin \(t-h\) or earlier.

We introduce the \textbf{multi-step conformal PID control} method
(hereafter MPID), which extends the PID method \citep{angelopoulos2024},
originally developed for one-step-ahead forecasting, to the multi-step
setting within the online learning framework of Section~\ref{sec-setup}.
For each forecast horizon \(h\in[H]\), define the P-control state
\begin{equation}\phantomsection\label{eq-p-state}{
q^{\mathrm P}_{t+h|t}=q^{\mathrm P}_{t+h-1|t-1}+\eta(\mathrm{err}_{t|t-h}-\alpha),
}\end{equation} with a finite initial state \(q^{\mathrm P}_{2h|h}\),
typically zero. The estimated \(1-\alpha\) quantile of the
\(h\)-step-ahead score at time \(t\) is updated as
\begin{equation}\phantomsection\label{eq-mpid}{
q_{t+h|t}=\underbrace{q^{\mathrm P}_{t+h|t}}_{\mathrm P}+\underbrace{r_t\Bigg(\sum_{i=h+1}^t(\mathrm{err}_{i|i-h}-\alpha)\Bigg)}_{\mathrm I}+\underbrace{\hat{s}_{t+h|t}}_{\mathrm D},
}\end{equation} where \(\eta > 0\) is a learning rate, which may vary
over time provided it is computed from information available at time
\(t\), and \(r_t\) is a saturation function that adheres to
\begin{equation}\phantomsection\label{eq-saturation_h}{
x \geq c \cdot g(t-h) \Longrightarrow r_t(x) \geq b, \quad \text {and} \quad x \leq-c \cdot g(t-h) \Longrightarrow r_t(x) \leq -b,
}\end{equation} for constants \(b, c > 0\) and an admissible function
\(g\) that is nonnegative, nondecreasing and sublinear, i.e.,
\(g(n)/n\to0\). For \(h=1\), Equation \eqref{eq-mpid} reduces to the
standard PID update. For each \(h\in[H]\), we apply this update
separately to the lower and upper tails, using the tail levels
\(\alpha^{-}\) and \(\alpha^{+}\) and the feedback available at time
\(t\), and combine the resulting thresholds via Equation
\eqref{eq-two-sided-pi} to form the prediction interval for \(y_{t+h}\).

The three components play distinct roles.

\emph{P (quantile tracking).} The P state moves up or down according to
whether \(y_t\) fell outside or inside the one-sided prediction set
issued at time \(t-h\). It thus incorporates the \(h\)-period feedback
delay explicitly: the most recent feedback usable at time \(t\) is
\(\mathrm{err}_{t|t-h}\).

\emph{I (error integration).} The I term accumulates the historical
coverage errors of the \(h\)-step-ahead one-sided prediction sets,
during updates, thereby enhancing the stability of coverage.

\emph{D (scorecasting).} The D term \(\hat{s}_{t+h|t}\) is a forecast of
the \(h\)-step-ahead nonconformity score, produced by any suitable
forecasting model (a ``scorecaster''; see Section~\ref{sec-acmcp})
trained on the \(h\)-step-ahead scores available up to and including
time \(t\). Its effectiveness depends on the ability of the scorecaster
to capture systematic and predictable structure in the scores, such as
temporal dependence or conditional heteroskedasticity. Only when such
structure is present and adequately modeled does the D term provide
additional benefits; otherwise, it can increase the variability of
coverage and of the prediction sets.

MPID is a special case of the generic update in Equation
\eqref{eq-acmcp_1}, with
\(\hat q_{t+h|t}=q^{\mathrm P}_{t+h|t}+\hat s_{t+h|t}\). Its one- and
two-sided prediction sets therefore inherit the finite-sample empirical
coverage bounds and long-run guarantees of Corollary~\ref{cor-cov_acmcp}
and Corollary~\ref{cor-cov-two-sided}, respectively, under the stated
assumptions. For reference, Appendix \ref{sec-extensions} presents
multi-step extensions of split conformal prediction \citep{vovk2005},
conformal prediction with weighted quantiles \citep{barber2023}, and ACI
\citep{gibbs2021}; we do not establish coverage guarantees for these.

\section{Autocorrelated multi-step conformal
prediction}\label{sec-acmcp}

MPID calibrates each horizon separately, and its scorecaster forecasts
the scores of a single horizon. In any non-trivial context, however,
multi-step forecast errors are correlated: the \(h\)-step-ahead error
\(e_{t+h|t}\) is correlated with the errors at the previous \(h-1\)
steps from the same origin, because errors propagate through the
recursive forecasting procedure. No existing conformal method exploits
this dependence.

This section studies the structure of multi-step forecast errors, uses
it to design the AcMCP scorecaster, and establishes coverage guarantees
that do not depend on how accurately the scorecaster approximates the
errors.

\subsection{Properties of multi-step forecast errors}\label{sec-ppt}

We assume that a time series \(\{y_t\}_{t \geq 1}\) follows a general
nonstationary autoregressive process:
\begin{equation}\phantomsection\label{eq-dgp}{
y_t = f_{t}(y_{(t-d):(t-1)}, \bm{x}_{(t-k):t}) + \varepsilon_t,
}\end{equation} where \(f_{t}\) depends on \(d\) lagged values of
\(y_t\) and on the current and \(k\) preceding values of the exogenous
predictors, and \(\varepsilon_t\) is the innovation. To isolate
uncertainty in future responses, we treat future exogenous predictors as
known. Let \(\mathcal F_t\) be the information generated by the
responses observed through time \(t\) and the exogenous predictor path.
The optimal point forecast under squared loss is
\(\hat y_{t+h|t}=\mathbb E(y_{t+h}|\mathcal F_t)\), with error
\(e_{t+h|t}=y_{t+h}-\hat y_{t+h|t}\).

At origin \(t\), \(p_h=\min\{d,h-1\}\) of the lagged responses entering
\(f_{t+h}\) are unobserved. For \(h\geq2\), collect them in
\(\bm z_{t,h}=(y_{t+h-1},\dots,y_{t+h-p_h})^{\prime}\), with conditional
mean \(\bm m_{t,h}=\mathbb E(\bm z_{t,h}|\mathcal F_t)\), and let
\(F_{t,h}\) denote \(f_{t+h}\) as a function of \(\bm z_{t,h}\) with all
other arguments held fixed. We write \(\partial_i\) for differentiation
in the \(i\)th coordinate, and \(\|\cdot\|_2\) and
\(\|\cdot\|_{\mathrm{op}}\) for the Euclidean and operator norms,
respectively.

\begin{proposition}[Autoregressive representation of optimal multi-step
forecast errors]\protect\hypertarget{prp-ar}{}\label{prp-ar}

Consider the process in Equation \eqref{eq-dgp}. Assume that each
\(f_s\) is twice continuously differentiable in the lagged responses,
that \(\mathbb E(y_s^2)<\infty\) and
\(\mathbb E(\varepsilon_s|\mathcal F_{s-1})=0\) for all \(s\geq1\), and
that the first-order Taylor terms and remainder below are integrable.
Then, for each \(h\in[H]\),
\begin{equation}\phantomsection\label{eq-ar}{
e_{t+h|t}
=\varepsilon_{t+h}+\sum_{i=1}^{p_h}\phi_{t,h,i}e_{t+h-i|t}+\widetilde R_{t,h},
}\end{equation} where \(\phi_{t,h,i}=\partial_iF_{t,h}(\bm m_{t,h})\)
and \(\widetilde R_{t,h}=R_{t,h}-\mathbb E(R_{t,h}|\mathcal F_t)\), with
\(R_{t,h}\) the remainder of the first-order Taylor expansion of
\(F_{t,h}\) at \(\bm m_{t,h}\). In particular,
\(\mathbb E(\widetilde R_{t,h}|\mathcal F_t)=0\) and
\(\widetilde R_{t,1}=0\).

Moreover, for \(h\geq2\), if there is a finite deterministic
\(M_{t,h}\geq0\) with \[
\sup_{u\in[0,1]}
\left\|\nabla^2F_{t,h}\!\left(\bm m_{t,h}+u(\bm z_{t,h}-\bm m_{t,h})\right)\right\|_{\mathrm{op}}
\leq M_{t,h}\quad\text{almost surely},
\] then \begin{equation}\phantomsection\label{eq-remainder-bound}{
\mathbb E(|\widetilde R_{t,h}|\,|\,\mathcal F_t)
\leq M_{t,h}\sum_{i=1}^{p_h}\mathbb E(e_{t+h-i|t}^2|\mathcal F_t).
}\end{equation}

\end{proposition}

\begin{proof}
See Appendix \ref{sec-proof_ar}.
\end{proof}

It is well-established in the forecasting literature that, for a
zero-mean covariance-stationary time series, the optimal linear
least-squares forecasts have \(h\)-step-ahead errors that are at most
MA\((h-1)\) process \citep{harvey1997}, a property that can be derived
using Wold's representation theorem. For the general nonstationary
autoregressive setting considered here, Proposition~\ref{prp-ma} gives
an innovation representation with an explicit remainder.

\begin{proposition}[Innovation representation of optimal multi-step
forecast errors]\protect\hypertarget{prp-ma}{}\label{prp-ma}

Under the assumptions of Proposition~\ref{prp-ar}, for each \(h\in[H]\),
\begin{equation}\phantomsection\label{eq-ma}{
e_{t+h|t}=\varepsilon_{t+h}+
\sum_{i=1}^{h-1}\theta_{t,h,i}\varepsilon_{t+h-i}+U_{t,h},
}\end{equation} where the coefficients \(\theta_{t,h,i}\) are
\(\mathcal F_t\)-measurable and the propagated remainders satisfy
\(U_{t,1}=0\) and
\(U_{t,j}=\widetilde R_{t,j}+\sum_{i=1}^{p_j}\phi_{t,j,i}U_{t,j-i}\) for
\(j\geq2\).

Moreover, for \(h\geq2\), suppose that the Hessian bounds of
Proposition~\ref{prp-ar} hold and that, for finite deterministic
constants \(M,L,V\geq0\) and all \(2\leq j\leq h\), \[
M_{t,j}\leq M,\qquad
\sum_{i=1}^{p_j}|\phi_{t,j,i}|\leq L,
\qquad \sum_{i=1}^{p_j}\mathbb E(e_{t+j-i|t}^2|\mathcal F_t)\leq V\quad\text{almost surely}.
\] Then \begin{equation}\phantomsection\label{eq-propagated-bound}{
\mathbb E\!\left(|U_{t,h}|\,|\,\mathcal F_t\right)
\leq MV\sum_{k=0}^{h-2}L^k\quad\text{almost surely}.
}\end{equation}

\end{proposition}

\begin{proof}
See Appendix \ref{sec-proof_ma}.
\end{proof}

Dropping the remainders in Propositions \ref{prp-ar} and \ref{prp-ma}
gives linear approximations that motivate regression and moving-average
specifications for multi-step errors, respectively. If the
autoregressive functions are linear in their lagged responses, allowing
intercepts, both remainders vanish and the representations are exact.
Otherwise, Equation \eqref{eq-remainder-bound} controls the Taylor
remainder through curvature and shorter-horizon mean squared errors, and
Equation \eqref{eq-propagated-bound} controls its propagation across
horizons.

Appendix \ref{sec-approximation-errors} shows that, with \(t\), \(h\),
and \(d\) fixed, both remainders have mean absolute size \(O(\delta^2)\)
and are therefore negligible relative to the innovation scale \(\delta\)
as \(\delta\downarrow0\). This requires deterministic bounds on
gradients and Hessians along the expansion segments that do not depend
on \(\delta\), and innovation conditional variances that are uniformly
\(O(\delta^2)\). The analysis motivates the AcMCP scorecaster in
Section~\ref{sec-novel}, a simplified specification that accounts for
dependence among forecast errors rather than reproducing their optimal
dynamics exactly.

\subsection{The AcMCP method}\label{sec-novel}

Motivated by these properties, we propose \textbf{autocorrelated
multi-step conformal prediction} (AcMCP). Whereas multi-step extensions
of existing conformal methods in Section~\ref{sec-mpid} treat horizons
separately, AcMCP exploits the dependence among multi-step forecast
errors. For each \(h\in[H]\), the \(h\)-step-ahead threshold is updated
online as \begin{equation}\phantomsection\label{eq-acmcp}{
q_{t+h|t}=q^{\mathrm P}_{t+h|t}+r_t\Bigg(\sum_{i=h+1}^t(\mathrm{err}_{i|i-h}-\alpha)\Bigg)+\tilde{s}_{t+h|t},
}\end{equation} where the lower- and upper-tail thresholds are updated
separately at levels \(\alpha^{-}\) and \(\alpha^{+}\) and combined into
a prediction interval via Equation \eqref{eq-two-sided-pi}. AcMCP thus
differs from MPID only in its scorecaster, which accounts for error
dependence.

Let \(\tilde e_{t+h|t}\) denote the forecast of \(e_{t+h|t}\), obtained
by equally combining two simple models: (i) an MA\((h-1)\) model fitted
to the \(h\)-step-ahead errors available up to and including time \(t\),
namely \(e_{1+h|1},\dots,e_{t|t-h}\), which captures serial dependence
within a horizon; and (ii) a linear regression of each \(h\)-step-ahead
error on the shorter-horizon errors from the same forecast origin, which
captures dependence across horizons and takes the recursively predicted
errors \(\tilde e_{t+h-1|t},\dots,\tilde e_{t+1|t}\) as inputs, since
the corresponding realized errors are not yet observed at time \(t\).
The score forecast in Equation \eqref{eq-acmcp} is
\(\tilde s_{t+h|t}=\tilde e_{t+h|t}\) for upper-tail calibration and
\(\tilde s_{t+h|t}=-\tilde e_{t+h|t}\) for lower-tail calibration.
Importantly, the role of \(\tilde e_{t+h|t}\) is not to forecast the
errors precisely, but to incorporate the autocorrelation structure of
multi-step forecast errors into the resulting prediction intervals.

The coverage guarantees below hold under the stated calibration
conditions and do not require the linear approximations in Propositions
\ref{prp-ar} and \ref{prp-ma} to be accurate.

\subsection{Coverage guarantees}\label{sec-cov}

Proposition~\ref{prp-cov_rt} and Corollaries
\ref{cor-cov_qt}, \ref{cor-cov_acmcp} concern single-tail calibration
with fixed target tail miscoverage level \(\alpha\).
Corollary~\ref{cor-cov-two-sided} combines the lower and upper tails at
levels \(\alpha^{-}\) and \(\alpha^{+}\) with
\(\alpha=\alpha^{-}+\alpha^{+}\) the total miscoverage level.

\begin{proposition}[]\protect\hypertarget{prp-cov_rt}{}\label{prp-cov_rt}

Fix \(h\in[H]\) and \(\alpha\in(0,1)\). Let \(\{s_{t+h|t}\}_{t\geq1}\)
be any sequence of finite scores in \((-b,b)\) with \(b\in(0,\infty]\),
and let \(r_t\) satisfy Equation \eqref{eq-saturation_h} and an
admissible function \(g\). Then the iteration
\(q_{t+h|t}=r_t\left(\sum_{i=h+1}^t(\mathrm{err}_{i|i-h}-\alpha)\right)\)
satisfies, for every \(T\geq h+1\),
\begin{equation}\phantomsection\label{eq-cov_rt}{
\left|\frac{1}{T-h}\sum_{t=h+1}^{T}(\mathrm{err}_{t|t-h}-\alpha)\right| \leq \frac{c \cdot g(T-h) + h}{T-h},
}\end{equation} with \(c\) the constant in Equation
\eqref{eq-saturation_h}. Consequently,
\(\lim_{T\to\infty}\frac{1}{T-h}\sum_{t=h+1}^{T}\mathrm{err}_{t|t-h}=\alpha\).

\end{proposition}

\begin{proof}
See Appendix \ref{sec-proof_cov_rt}.
\end{proof}

For \(h=1\), Proposition~\ref{prp-cov_rt} recovers Proposition 2 of
\citet{angelopoulos2024}. With \(n=T-h\) evaluated forecasts, the bound
decomposes into \(c\,g(n)/n\), determined by the integrator's
saturation, and \(h/n\), which captures the feedback delay. For fixed
\(n\), a longer horizon enlarges the delay term, although actual
calibration errors need not grow. For every fixed \(h\) the bound
vanishes because \(g\) is sublinear, so the delay affects the
finite-sample guarantee but not long-run calibration.

The constant-step quantile iteration is a special case of
Proposition~\ref{prp-cov_rt}, so its calibration bound follows directly.

\begin{corollary}[]\protect\hypertarget{cor-cov_qt}{}\label{cor-cov_qt}

Fix \(h\in[H]\) and \(\alpha\in(0,1)\). Let \(\{s_{t+h|t}\}_{t\geq1}\)
be any sequence of finite scores in \((-b,b)\) with \(0<b<\infty\), and
let \(q_{2h|h}\in\mathbb R\) be an arbitrary initial threshold. Then the
iteration \(q_{t+h|t}=q_{t+h-1|t-1}+\eta(\mathrm{err}_{t|t-h}-\alpha)\)
with \(\eta>0\) satisfies, for every \(T\geq h+1\), \[
\left|\frac{1}{T-h}\sum_{t=h+1}^{T}(\mathrm{err}_{t|t-h}-\alpha)\right| \leq \frac{b+|q_{2h|h}|+\eta h}{\eta(T-h)}.
\] In particular,
\(\lim_{T\to\infty}\frac{1}{T-h}\sum_{t=h+1}^{T}\mathrm{err}_{t|t-h}=\alpha\).

\end{corollary}

\begin{proof}
See Appendix \ref{sec-proof_cov_qt}.
\end{proof}

For \(h=1\), Corollary~\ref{cor-cov_qt} recovers Proposition 1 of
\citet{angelopoulos2024}. The multi-step bound,
\((b+|q_{2h|h}|)/(\eta n)+h/n\), separates the effects of score
magnitude, initialization, and learning rate from that of the feedback
delay. The boundedness of the scores is a purely theoretical assumption
that keeps the accumulated miscoverage error of the quantile-tracking
recursion under control. In practice, the bound itself need not be
known, and the update runs unchanged as long as the scores are finite.

More importantly, Proposition~\ref{prp-cov_rt} also establishes
one-sided calibration for AcMCP. We first write Equation
\eqref{eq-acmcp} as \begin{equation}\phantomsection\label{eq-acmcp_1}{
q_{t+h|t}=\hat{q}_{t+h|t}+r_t\left(\sum_{i=h+1}^t (\mathrm{err}_{i|i-h}-\alpha)\right),
}\end{equation} where \(\hat{q}_{t+h|t}\) may be any function of the
past observations \(\{(\bm{x}_i,y_i)\}_{1\leq i\leq t}\) and the earlier
thresholds \(q_{i+h|i}\), \(i\leq t-1\), and can be viewed as a forecast
of the next threshold from the available history. Taking
\(\hat q_{t+h|t}=q^{\mathrm P}_{t+h|t}+\tilde s_{t+h|t}\) recovers
AcMCP, and taking
\(\hat q_{t+h|t}=q^{\mathrm P}_{t+h|t}+\hat s_{t+h|t}\) recovers MPID.

\begin{corollary}[]\protect\hypertarget{cor-cov_acmcp}{}\label{cor-cov_acmcp}

Fix \(h\in[H]\) and \(\alpha\in(0,1)\). Let the scores \(s_{t+h|t}\) and
baselines \(\hat q_{t+h|t}\) be finite with
\(|s_{t+h|t}-\hat q_{t+h|t}|<b\) for all \(t\geq1\), where
\(b\in(0,\infty]\), and let \(r_t\) satisfy Equation
\eqref{eq-saturation_h} and an admissible function \(g\). Then the
iteration in Equation \eqref{eq-acmcp_1} satisfies, for every
\(T\geq h+1\), \[
\left|\frac{1}{T-h}\sum_{t=h+1}^{T}(\mathrm{err}_{t|t-h}-\alpha)\right|
\leq \frac{c \cdot g(T-h)+h}{T-h}.
\] In particular,
\(\lim_{T\to\infty}\frac{1}{T-h}\sum_{t=h+1}^{T}\mathrm{err}_{t|t-h}=\alpha\).

\end{corollary}

\begin{proof}
See Appendix \ref{sec-proof_cov_acmcp}.
\end{proof}

Under the stated conditions, integral feedback ensures calibration for
any baseline \(\hat q_{t+h|t}\), and thus without requiring accurate
forecasts from the scorecaster. The scorecaster instead aims to improve
interval efficiency by accounting for error dependence, although this
improvement is not guaranteed by the corollary. Unlike
Corollary~\ref{cor-cov_qt}, which bounds the scores themselves,
Corollary~\ref{cor-cov_acmcp} bounds only their deviations from the
baselines, \(s_{t+h|t}-\hat q_{t+h|t}\).

The next corollary combines the lower- and upper-tail calibration
guarantees to establish coverage for the asymmetric prediction interval
in Equation \eqref{eq-two-sided-pi}.

\begin{corollary}[]\protect\hypertarget{cor-cov-two-sided}{}\label{cor-cov-two-sided}

Fix \(h\in[H]\) and \(\alpha\in(0,1)\). Construct the interval in
Equation \eqref{eq-two-sided-pi} from thresholds obtained by applying
Equation \eqref{eq-acmcp_1} separately to the lower- and upper-tail
scores, with targets \(\alpha^{-},\alpha^{+}>0\) satisfying
\(\alpha^{-}+\alpha^{+}=\alpha\). Suppose each tail satisfies the
assumptions of Corollary~\ref{cor-cov_acmcp}. Then, for every
\(T\geq h+1\),
\begin{equation}\phantomsection\label{eq-two-sided-finite}{
\frac{1}{T-h}\sum_{t=h+1}^{T}\mathbbm{1}\{y_t\in\hat{\mathcal C}_{t|t-h}^{\alpha}\}
\geq 1-\alpha-2\frac{c\cdot g(T-h)+h}{T-h},
}\end{equation} and hence
\(\liminf_{T\to\infty}\frac{1}{T-h}\sum_{t=h+1}^{T}\mathbbm{1}\{y_t\in\hat{\mathcal C}_{t|t-h}^{\alpha}\}\geq1-\alpha\).

Furthermore, let
\(\chi_{T,h}:=\frac{1}{T-h}\sum_{t=h+1}^{T}\mathbbm{1}\{\hat y_{t|t-h}-q_{t|t-h}^{-}>\hat y_{t|t-h}+q_{t|t-h}^{+}\}\)
be the frequency of crossed endpoints, that is, of empty intervals. Then
\begin{equation}\phantomsection\label{eq-two-sided-enhanced}{
\left|\frac{1}{T-h}\sum_{t=h+1}^{T}\mathbbm{1}\{y_t\in\hat{\mathcal C}_{t|t-h}^{\alpha}\}
-(1-\alpha)\right|\leq 2\frac{c \cdot g(T-h)+h}{T-h}+\chi_{T,h},
}\end{equation} so the empirical coverage converges to \(1-\alpha\)
whenever \(\chi_{T,h}\to0\).

\end{corollary}

\begin{proof}
See Appendix \ref{sec-proof_cov_two_sided}.
\end{proof}

Unlike the preceding single-tail results,
Corollary~\ref{cor-cov-two-sided} provides a lower bound on empirical
coverage of intervals and ensures a limit inferior of at least
\(1-\alpha\), while convergence to exactly \(1-\alpha\) follows when the
endpoint-crossing frequency vanishes. Setting \(\hat q_{t+h|t}=0\), the
result also covers the integral-only iteration in
Proposition~\ref{prp-cov_rt}. Analogous two-sided guarantees for the
constant-step quantile-tracking update shown in
Corollary~\ref{cor-cov_qt} are given in
Corollary~\ref{cor-cov-two-sided-qt} in Appendix
\ref{sec-two-sided-constant-step}.

These results bound empirical coverage at each forecast horizon, both in
finite samples and in the long run, but do not ensure conditional
coverage at individual forecast origins or simultaneous coverage across
horizons. No method can provide such conditional guarantees in finite
samples without distributional assumptions: even for i.i.d. data, a
procedure with distribution-free conditional coverage must produce
intervals of infinite expected length whenever the covariate
distribution is nonatomic \citep{lei2014, barber2021limits}, and our
setting includes i.i.d. data as a special case. Nor do they guarantee
nominal coverage in every local window. Appendix
\ref{sec-local-coverage} shows that the coverage shortfall over a window
is bounded by twice the sum of the cumulative calibration bounds at its
endpoints, divided by the window length. This bound vanishes when the
window length grows faster than that sum, but may remain uninformative
for fixed-length rolling windows. Thus, long-run calibration alone does
not guarantee protection against local undercoverage.

\section{Experiments}\label{sec-experiments}

We evaluate MPID and AcMCP using two simulated settings and two
real-world data sets. For benchmarks, we use the multi-step extensions
of split conformal prediction (MSCP), weighted conformal prediction
(MWCP), and adaptive conformal prediction (MACP) described in Appendix
\ref{sec-extensions}, together with the multi-step sequential predictive
conformal inference (SPCI) proposed by \citet{xu2023}. The following
parameter choices are used:

\begin{enumerate}
\def\labelenumi{(\alph{enumi})}
\tightlist
\item
  we target coverage \(1-\alpha=0.9\), with equal tail miscoverage
  levels \(\alpha^{-}=\alpha^{+}=0.05\);
\item
  for MWCP, we use \(b=0.99\) as per \citet{barber2023};
\item
  following \citet{angelopoulos2024}, MACP uses a step size
  \(\gamma=0.005\); MPID uses a Theta model as the scorecaster; and both
  MPID and AcMCP use a learning rate of \(\eta=0.1\hat{B}_t\) for
  quantile tracking, where
  \(\hat{B}_t=\max\{s_{t-\Delta+1|t-\Delta-h+1}, \dots, s_{t|t-h}\}\) is
  the highest score over a trailing window of length \(\Delta\), which
  is set equal to the calibration set size. Because this learning rate
  varies over time, Corollary~\ref{cor-cov_qt}, which assumes a constant
  step, does not apply; however, \(\hat{B}_t\) depends only on past
  scores, so the P state remains a function of the past and the
  guarantees of Corollaries \ref{cor-cov_acmcp} and
  \ref{cor-cov-two-sided} continue to hold;
\item
  we adopt a nonlinear saturation function
  \(r_t(x)=K_{\mathrm I} \tan \left(x \log (t) / (t C_{\text {sat }})\right)\),
  where \(\tan (x)=\operatorname{sign}(x) \cdot \infty\) for
  \(x \notin[-\pi / 2, \pi / 2]\), and constants
  \(C_{\text {sat }}, K_{\mathrm{I}}>0\) are chosen heuristically as in
  \citet{angelopoulos2024};
\item
  we use a clipped version of MACP, replacing an infinite upper (lower)
  bound by the point forecast plus (minus) the largest upper-tail
  (lower-tail) score observed by the forecast origin;
\item
  for SPCI, we follow the implementation of \citet{xu2023}, using a
  conditional quantile forest with \(10\) trees and maximum depth \(2\).
  For each horizon, we refit it over the rolling calibration window
  using the latest \(10\) available signed errors as predictors and
  construct intervals at fixed quantile levels \(0.05\) and \(0.95\).
  All methods share point forecasts and out-of-sample calibration
  errors.
\end{enumerate}

\subsection{Simulated linear autoregressive process}\label{sec-simar}

We first consider a simulated stationary time series generated from a
simple AR\((2)\) process \[
y_t = 0.8y_{t-1} - 0.5y_{t-2} + \varepsilon_t,
\] where \(\varepsilon_t\) is white noise with error variance
\(\sigma^2 = 1\). After an appropriate burn-in period, we generate
\(N=5000\) data points. Under the sequential split and online learning
settings, we create training sets \(\mathcal{D}_{\text{tr}}\) and
calibration sets \(\mathcal{D}_{\text{cal}}\), each with a length of
\(500\). We use AR\((2)\) models to generate \(1\)- to \(3\)-step-ahead
point forecasts (i.e.~\(H=3\)), using the \texttt{Arima()} function from
the \texttt{forecast} R package \citep{hyndman2008, hyndman2024}. The
goal is to generate prediction intervals using various proposed
conformal prediction methods and evaluate whether they can achieve the
nominal long-run coverage for each separate forecast horizon.

\begin{figure}

\centering{

\pandocbounded{\includegraphics[keepaspectratio]{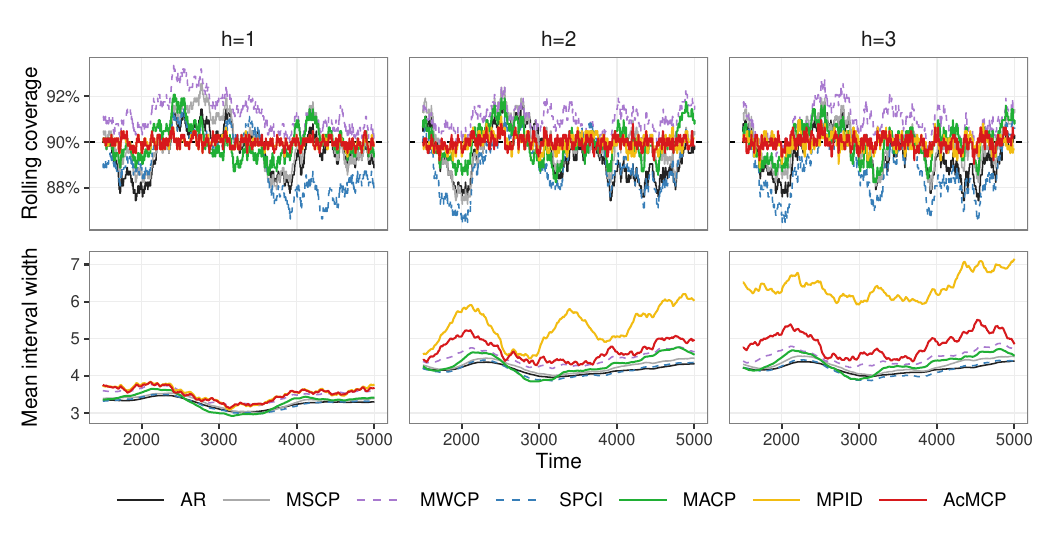}}

}

\caption{\label{fig-AR2_cov}AR(2) simulation results showing rolling
coverage (top) and mean interval width (bottom) for each forecast
horizon. Both metrics are computed over rolling windows of size \(500\).
The black dashed horizontal line indicates the target coverage level
\(1-\alpha=0.9\).}

\end{figure}%

\begin{figure}

\centering{

\pandocbounded{\includegraphics[keepaspectratio]{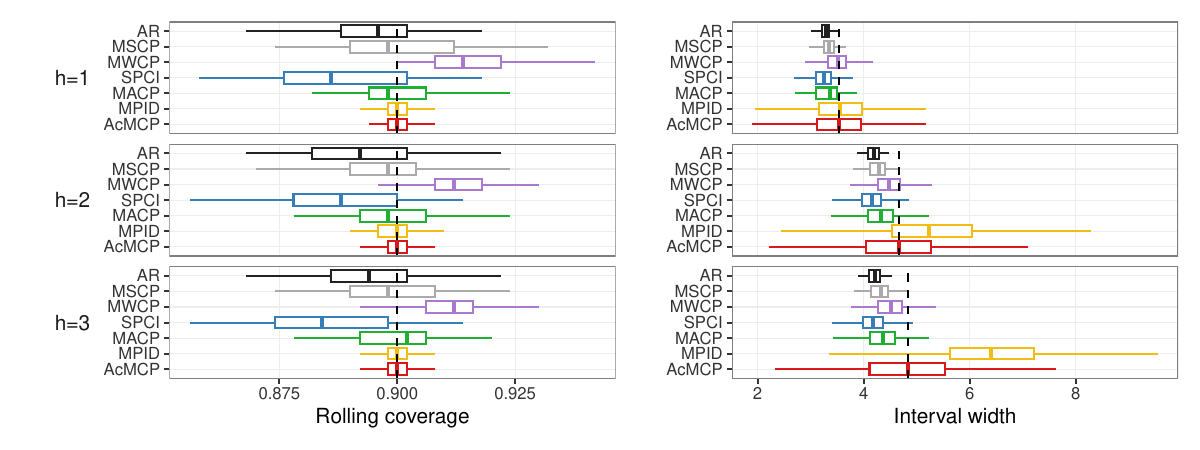}}

}

\caption{\label{fig-AR2_box}AR(2) simulation results showing boxplots of
the rolling coverage and interval width for each method across different
forecast horizons. Black dashed lines indicate the target coverage level
in the left panels and the median interval width of AcMCP in the right
panels.}

\end{figure}

Figure~\ref{fig-AR2_cov} presents the rolling coverage and mean interval
width of each method for each forecast horizon, computed over rolling
windows of size \(500\); rolling median widths behave similarly and are
omitted. The analytic intervals from the fitted AR model are also shown.
MPID and AcMCP stay close to the \(90\%\) target across windows, whereas
the other methods, including the AR model, swing much further from it.
AcMCP adapts its interval widths to past coverage, widening them when
competing methods undercover. It attains coverage similar to MPID with
narrower intervals, an advantage that grows with \(h\) without
deteriorating at smaller \(h\). Figure~\ref{fig-AR2_box} presents the
corresponding boxplots of rolling coverage and interval width.

\begin{figure}[!b]

\centering{

\pandocbounded{\includegraphics[keepaspectratio]{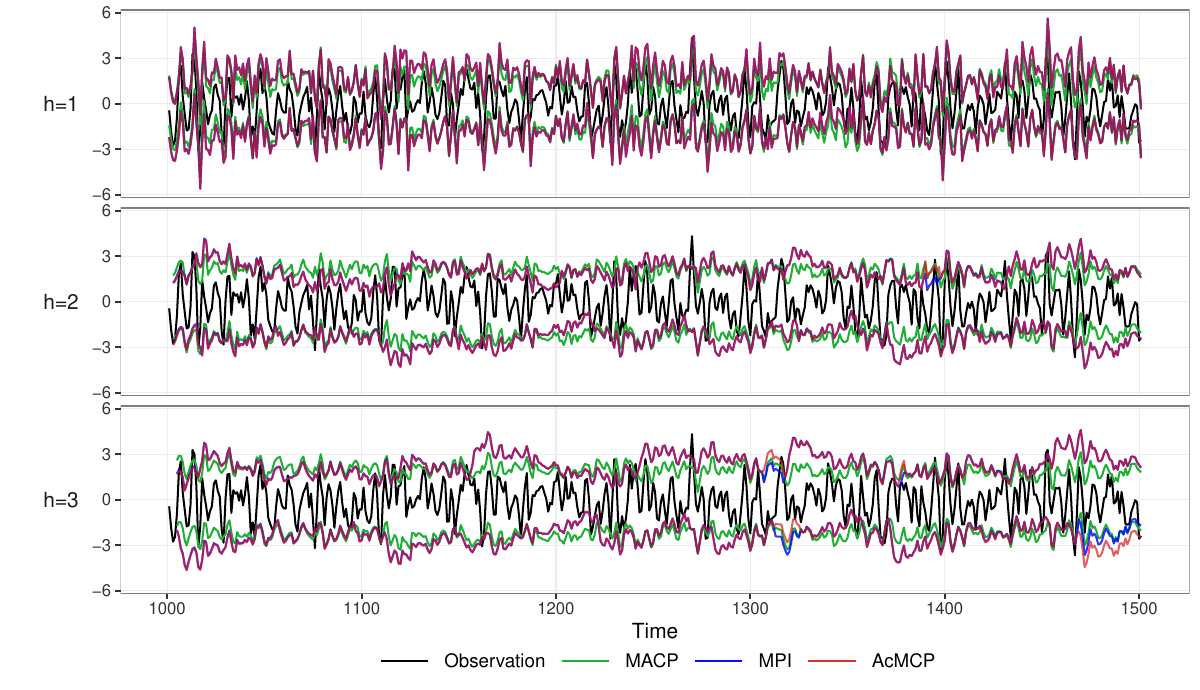}}

}

\caption{\label{fig-AR2_timeplot}AR(2) simulation results showing the
prediction interval bounds for the MACP, MPI, and AcMCP methods over a
truncated period of length 500.}

\end{figure}%

To examine the effect of the scorecasting term \(\tilde s_{t+h|t}\), we
compare AcMCP with the version without this term, denoted by MPI.
Figure~\ref{fig-AR2_timeplot} displays their prediction intervals over a
selected period of \(500\) observations. The two sets of bounds mostly
overlap; the main differences occur around time 1320 and during
1470--1500, where AcMCP fans out locally relative to MPI, most visibly
at \(h=3\). The figure also shows MACP, which likewise updates its
intervals using past miscoverage. AcMCP and MACP both capture some
patterns in the observations, and neither consistently yields narrower
intervals.

\subsection{Simulated nonlinear autoregressive
process}\label{simulated-nonlinear-autoregressive-process}

Next, consider the case of a nonlinear data-generating process, defined
as \[
y_t = \sin(y_{t-1}) + 0.5\log(y_{t-2} + 1) + 0.1y_{t-1}x_{1,t} + 0.3x_{2,t} + \varepsilon_{t},
\] where \(x_{1,t}\) and \(x_{2,t}\) are uniformly distributed on
\([0,1]\), and \(\varepsilon_{t}\) is white noise with variance
\(\sigma^2=0.01\). Thus, \(y_t\) nonlinearly depends on its lagged
values \(y_{t-1}\), \(y_{t-2}\), and exogenous variables \(x_{1,t}\) and
\(x_{2,t}\).

\begin{figure}[!b]

\centering{

\pandocbounded{\includegraphics[keepaspectratio]{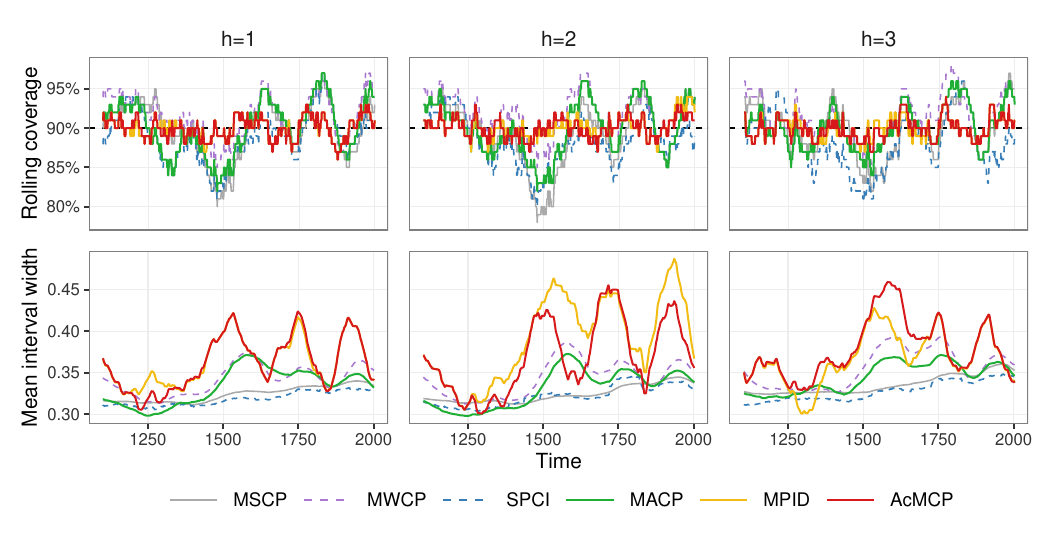}}

}

\caption{\label{fig-NL_cov}Nonlinear simulation results showing rolling
coverage (top) and mean interval width (bottom) for each forecast
horizon. Both metrics are computed over rolling windows of size \(100\).
The black dashed horizontal line indicates the target coverage level
\(1-\alpha=0.9\).}

\end{figure}%

\begin{figure}[!b]

\centering{

\pandocbounded{\includegraphics[keepaspectratio]{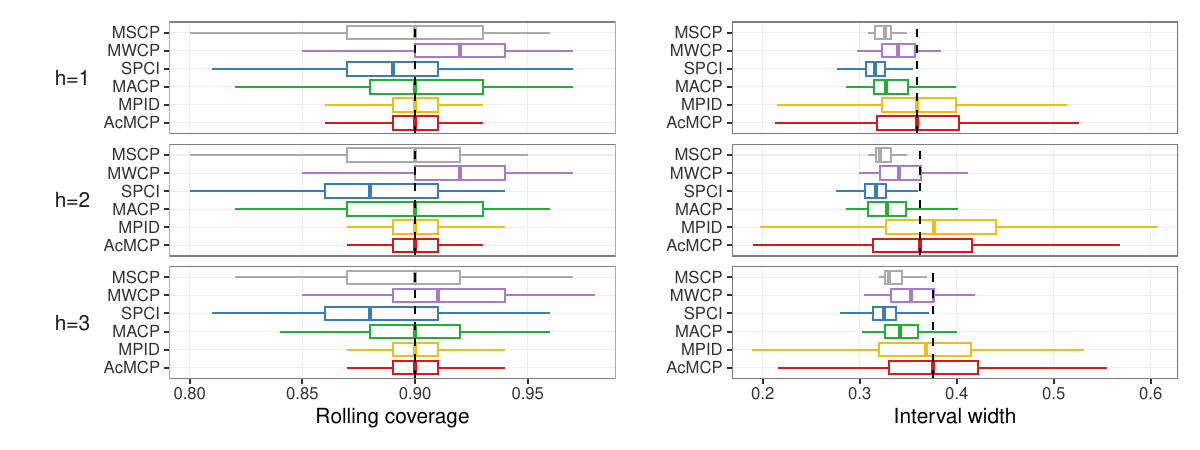}}

}

\caption{\label{fig-NL_box}Nonlinear simulation results showing boxplots
of the rolling coverage and interval width for each method across
different forecast horizons. Black dashed lines indicate the target
coverage level in the left panels and the median interval width of AcMCP
in the right panels.}

\end{figure}%

After an appropriate burn-in period, we generate \(N=2000\) data points.
Under the sequential split and online learning settings, both the
training set \(\mathcal{D}_{\text{tr}}\) and the calibration set
\(\mathcal{D}_{\text{cal}}\) have size \(500\). Given the nonlinear DGP,
we use feed-forward neural networks with a single hidden layer and
lagged inputs to generate \(1\)- to \(3\)-step-ahead point forecasts
(\(H=3\)), using the \texttt{nnetar()} function from the
\texttt{forecast} R package \citep{hyndman2024}. Note that analytic
prediction intervals are unavailable for neural networks, so we do not
include them when presenting the results.

Figure~\ref{fig-NL_cov} shows the rolling coverage and mean interval
width of each method over rolling windows of size \(100\). MPID and
AcMCP fluctuate only slightly around the \(90\%\) target, whereas MSCP,
MWCP, SPCI and MACP show pronounced fluctuations. To maintain coverage,
MPID and AcMCP vary their interval widths more than the competing
methods. AcMCP generally produces narrower intervals than MPID at
\(h=2\) but wider intervals at \(h=3\). Under AcMCP, the interval width
at \(h=3\) exceeds that at \(h=2\) in \(53.41\%\) of the test cases,
compared with \(38.08\%\) under MPID, which is more consistent with
forecast uncertainty increasing with the horizon. The boxplots in
Figure~\ref{fig-NL_box} confirm that MPID and AcMCP have less variable
rolling coverage than MSCP, MWCP, SPCI and MACP, but larger and more
variable interval widths.

\subsection{Electricity demand data}\label{electricity-demand-data}

\begin{figure}[!hb]

\centering{

\pandocbounded{\includegraphics[keepaspectratio]{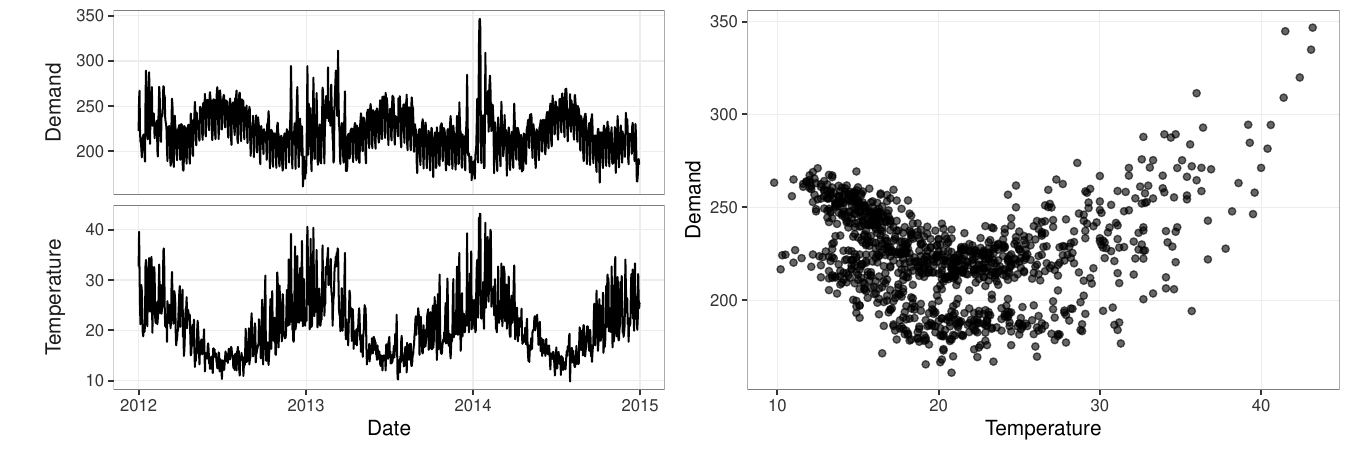}}

}

\caption{\label{fig-elec_data}Daily electricity demand and corresponding
daily maximum temperatures in 2012--2014, Victoria, Australia.}

\end{figure}

We apply the conformal prediction methods using data comprising daily
electricity demand (GW), daily maximum temperature (degrees Celsius),
and holiday information for Victoria, Australia, from 2012 to 2014.
Temperatures are taken from the Melbourne Regional Office weather
station. The left panel of Figure~\ref{fig-elec_data} displays the daily
electricity demand and temperatures over the period. The right panel
shows a nonlinear relationship between demand and temperature, with
demand increasing for low temperatures (due to heating) and increasing
for high temperatures (due to cooling); the two clouds of points
correspond to working days and non-working days.

The response variable is \(\texttt{Demand}\), with two covariates:
\(\texttt{Temperature}\) and \(\texttt{Workday}\) (an indicator variable
for working days). Following \citet{hyndman2021}, we fit a dynamic
regression model with a piecewise linear function of temperature (with a
knot at \(18\) degrees) to generate \(1\)- to \(7\)-step-ahead point
forecasts (\(H=7\)), modeling the errors with an ARIMA process to
capture autocorrelation. We use two years of data for training and
\(100\) observations for calibration.

\begin{figure}[!hbtp]

\centering{

\pandocbounded{\includegraphics[keepaspectratio]{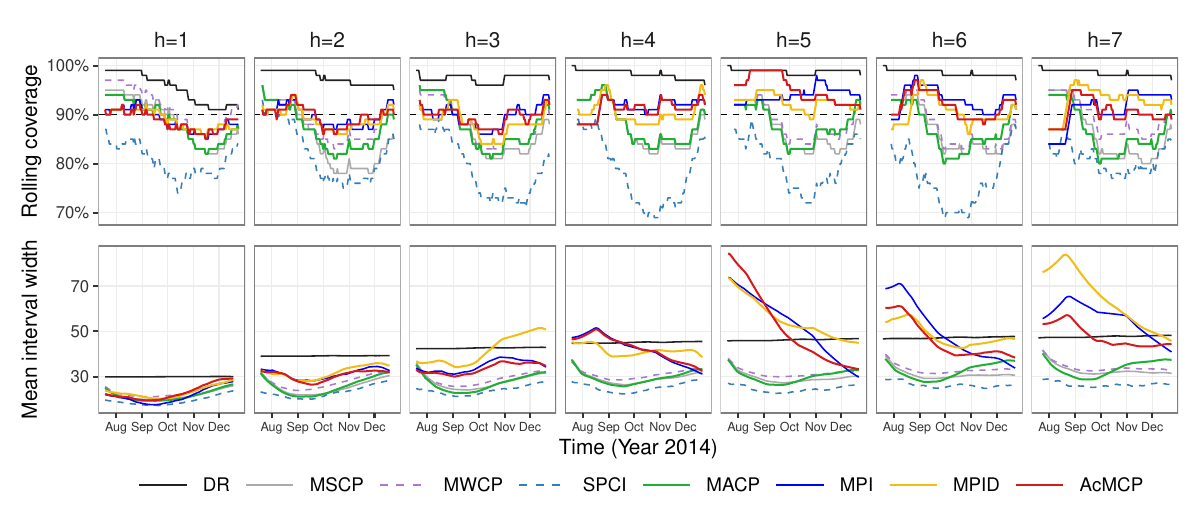}}

}

\caption{\label{fig-elec_cov}Electricity demand data results showing
rolling coverage (top) and mean interval width (bottom) for each
forecast horizon. Both metrics are computed over rolling windows of size
\(100\). The black dashed horizontal line indicates the target coverage
level \(1-\alpha=0.9\).}

\end{figure}

\begin{figure}[!hb]

\centering{

\pandocbounded{\includegraphics[keepaspectratio]{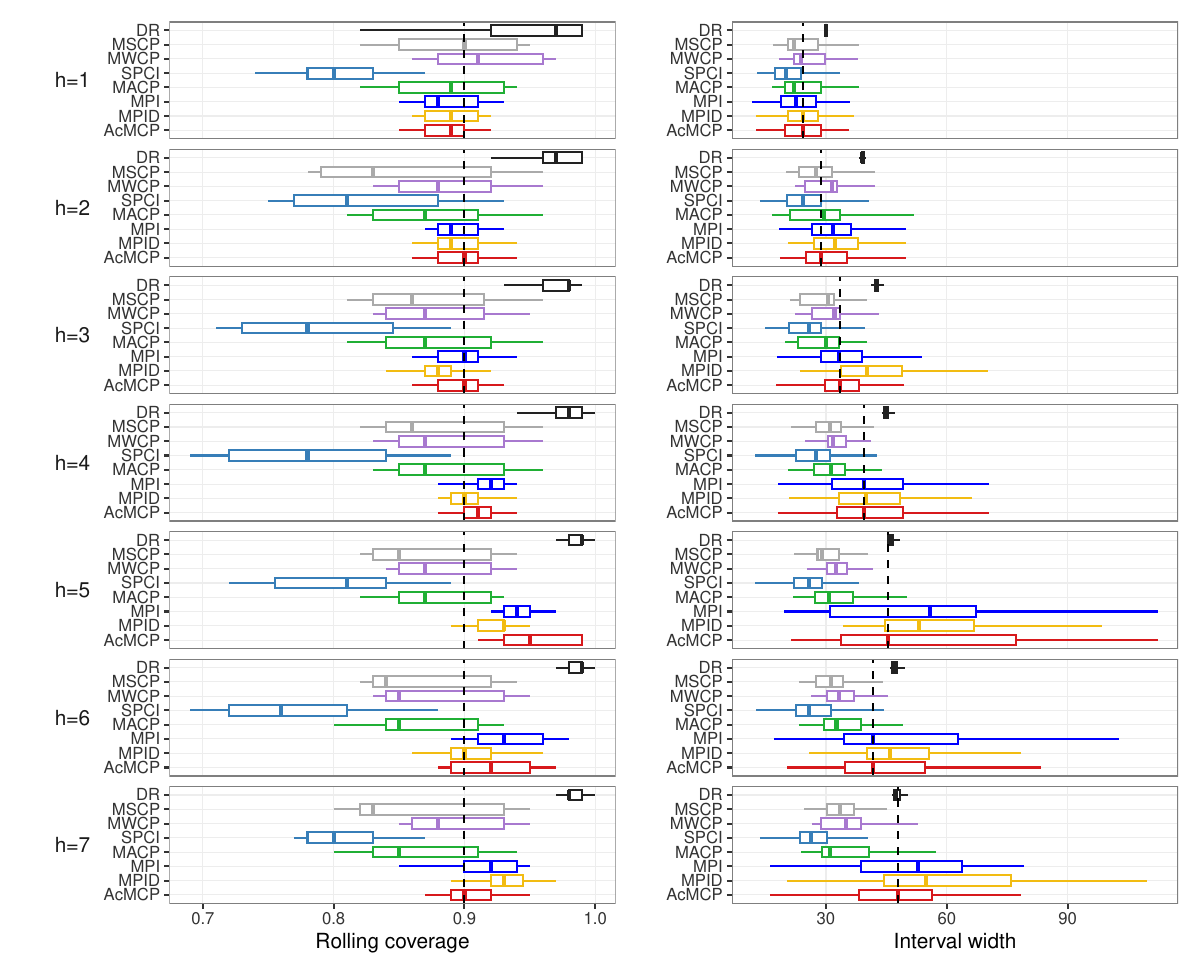}}

}

\caption{\label{fig-elec_box}Electricity demand data results showing
boxplots of the rolling coverage and interval width for each method
across different forecast horizons. Black dashed lines indicate the
target coverage level in the left panels and the median interval width
of AcMCP in the right panels.}

\end{figure}

Figure~\ref{fig-elec_cov} and Figure~\ref{fig-elec_box} compare rolling
coverage and interval width across methods over rolling windows of size
\(100\); DR denotes the analytic intervals from the dynamic regression
model. DR consistently overcovers, giving much wider intervals than the
other methods for \(h\leq4\). MSCP, MWCP and MACP fluctuate
substantially in rolling coverage and undercover in many windows after
September 2014 at all horizons. SPCI produces narrower intervals than
MPID and AcMCP, but at the cost of substantial undercoverage across
horizons. MPI, MPID and AcMCP give wider intervals that largely avoid
the post-September 2014 undercoverage; among these, MPID has slightly
poorer coverage than MPI and AcMCP at \(h=3\) despite wider intervals.
AcMCP has coverage similar to MPI but narrower intervals, particularly
at longer horizons, and smaller mean interval widths than MPID at every
horizon, with reductions of \(18.94\%\) at \(h=3\) and \(22.29\%\) at
\(h=7\).

AcMCP achieves overall coverage close to the nominal level at most
horizons, although it overcovers at \(h=5\), and has smaller mean and
median interval widths than MPID at every horizon. These properties
support its use in short-term generation planning: combined with supply
forecasts, the intervals can flag days that warrant closer review of
capacity margins and support assessments of candidate generation plans
under different demand scenarios. However, satisfactory overall and
rolling coverage does not imply conditional coverage under extreme
temperatures, high demand, or other critical operating conditions. The
theoretical guarantees concern empirical coverage at each horizon; they
ensure neither coverage conditional on the information available for a
given decision nor simultaneous coverage across horizons. Costly
upper-tail exceedances can therefore occur despite satisfactory average
coverage. Before using the intervals to set reserves, practitioners
should assess the frequency and magnitude of upper-bound exceedances
during high-demand periods, and weigh this exposure against available
generation capacity and the cost of covering any shortfall.

\subsection{Eating out expenditure
data}\label{eating-out-expenditure-data}

\begin{figure}[!tb]

\centering{

\pandocbounded{\includegraphics[keepaspectratio]{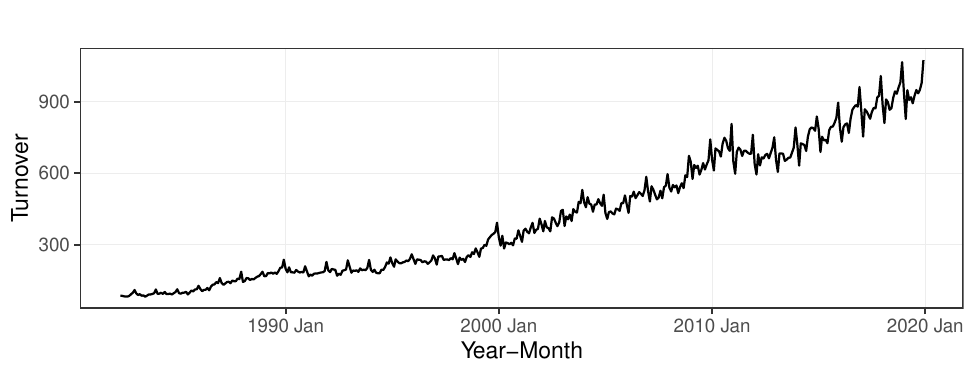}}

}

\caption{\label{fig-cafe_data}Monthly expenditure on cafes, restaurants
and takeaway food services in Victoria, Australia, from April 1982 to
December 2019.}

\end{figure}

In our final example, we apply the conformal prediction methods to
forecast the eating out expenditure (in millions of dollars) in
Victoria, Australia. The data include monthly expenditure on cafes,
restaurants and takeaway food services from April 1982 to December 2019,
as shown in Figure~\ref{fig-cafe_data}. The data show an overall upward
trend, clear annual seasonality, and level-dependent variability.

We consider three models: ARIMA with logarithmic transformation, ETS,
and STL-ETS \citep{hyndman2021}, and then output their simple average as
final point forecasts. STL-ETS applies ETS to the seasonally adjusted
series obtained via STL decomposition. All models are automatically
trained using the \texttt{forecast} R package \citep{hyndman2024}. We
aim to forecast \(12\) months ahead (\(H=12\)) using \(20\) years of
data for training and \(5\) years for calibration. The test period
contains \(153\) months for one-step evaluation, with fewer observed
targets at longer horizons. We therefore summarize coverage and interval
width over the available evaluation periods rather than showing rolling
windows.

\begin{figure}

\centering{

\pandocbounded{\includegraphics[keepaspectratio]{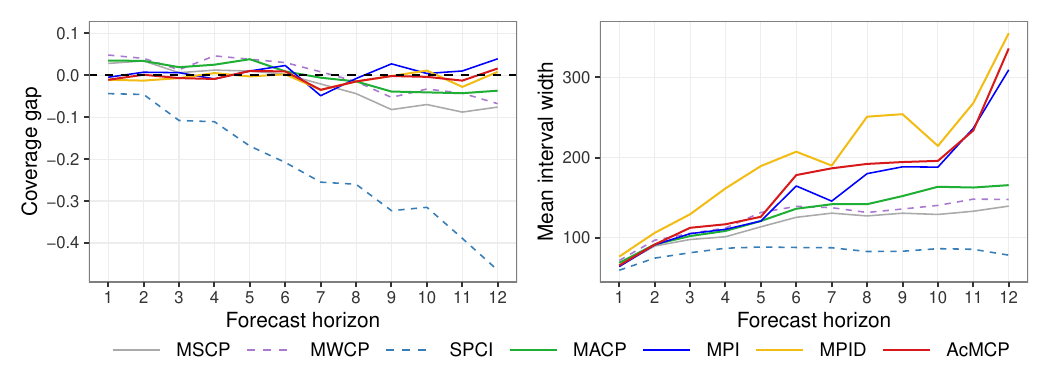}}

}

\caption{\label{fig-cafe_cov}Eating out expenditure data results showing
empirical coverage minus the \(90\%\) target level (left) and mean
interval width (right), computed over the entire test set for each
forecast horizon. The black dashed line indicates zero coverage gap.}

\end{figure}

Because the horizons in this application are relatively long,
Figure~\ref{fig-cafe_cov} reports the coverage gap and mean interval
width over the whole test set for each method and horizon. SPCI produces
narrower intervals but undercovers at every horizon, especially longer
ones. MSCP, MWCP and MACP are conservative at short horizons,
particularly MWCP and MACP, but all three undercover for \(h\geq8\);
overall, the coverage of MSCP, MWCP, SPCI and MACP declines as the
horizon increases. MPI and AcMCP approximately achieve the target
coverage for \(h\leq5\) with mean widths comparable to those of MSCP,
MWCP and MACP, and keep coverage within a tighter range across horizons,
albeit at the cost of interval efficiency. AcMCP has smaller mean
interval widths than MPID at all twelve horizons.

\subsection{Post-hoc comparison}\label{post-hoc-comparison}

For MACP, the step size \(\gamma\) controls both the calibration
tightness and the responsiveness of the interval widths: smaller values
lead to smoother interval evolution but slower correction of local
coverage errors, while larger values improve responsiveness at the cost
of increased variability. An analogous trade-off arises in AcMCP, where
the learning rate \(\eta\) governs the balance between adaptivity and
stability. This motivates a comparison with MACP tuned for comparable
local coverage variability.

To examine this, we conduct a post-hoc comparison in which the
learning-rate rule and tuning constants for AcMCP are held fixed, while
\(\gamma\) for MACP is selected by grid search over \([0.005, 0.5]\) in
steps of \(0.015\). Specifically, we choose the smallest \(\gamma\) for
which the sample variance of one-step-ahead local coverage over rolling
windows of the test set does not exceed that of AcMCP, and use it at all
horizons. This is a post-hoc comparison based on the test set, which
favors MACP, rather than a prospective tuning procedure or a guarantee
of optimality. The selected step sizes, \(\gamma=0.065\), \(0.05\),
\(0.035\) and \(0.035\) for the four data sets, are much larger than the
default \(\gamma=0.005\). The working miscoverage level of MACP
therefore often falls low enough to produce infinite bounds, which are
clipped as described in Section~\ref{sec-experiments}; depending on the
data set and horizon, approximately \(12\%\) to \(89\%\) of the issued
tuned MACP intervals have at least one bound clipped.

Table~\ref{tbl-posthoc} summarizes, for each forecast horizon, mean
coverage on observed targets, the minimum and maximum local coverage
across rolling windows, and the mean and median widths of issued
intervals for MACP and AcMCP. In the AR(2) simulation, both methods
attain mean coverage close to the nominal \(90\%\) with similar ranges
of local coverage, and AcMCP produces narrower intervals at every
horizon. In the nonlinear simulation, both methods achieve near-nominal
coverage. Neither method consistently yields smaller mean interval
widths across horizons, whereas MACP consistently has slightly smaller
median widths.

For electricity demand, tuned MACP achieves near-nominal overall
coverage but exhibits local undercoverage, with minimum local coverage
as low as \(84\%\). AcMCP has higher minimum local coverage at most
horizons, although it overcovers at \(h=5\). AcMCP also has smaller mean
interval widths at six horizons, with reductions of up to approximately
\(19\%\). MACP has smaller median widths at most horizons, but
occasional very wide intervals inflate its mean widths even after
clipping. For eating out expenditure, the clearest differences arise at
longer horizons. AcMCP has overall coverage closer to the target for
\(h\geq9\) and higher minimum local coverage for \(h\geq8\), with the
latter improving by approximately \(10\)--\(18.3\) percentage points.
However, AcMCP intervals are considerably wider for \(h\geq6\). This
additional width is not always accompanied by better overall coverage:
at \(h=6\) and \(h=8\), the two methods have similar overall coverage,
while at \(h=7\) MACP attains higher coverage with narrower intervals.
Overall, tuning the step size substantially narrows the gap between MACP
and AcMCP. AcMCP, which is not tuned on the test set, retains more
uniform coverage across horizons, but its intervals are not uniformly
shorter: at longer horizons in the real data sets, they can be wider
than those of a tuned MACP with similar coverage.

\begin{table}

\caption{\label{tbl-posthoc}Performance of MACP and AcMCP across the four datasets, summarized by mean test-set coverage, minimum and maximum local coverage over rolling windows, and mean and median interval widths for each forecast horizon.}

\centering{

\resizebox{\linewidth}{!}{
\begin{tabular}{lrrrrrrrrrr}
\toprule
     & \multicolumn{2}{c}{Mean coverage (\%)}               & \multicolumn{2}{c}{Min. local coverage (\%)}              & \multicolumn{2}{c}{Max. local coverage (\%)}              & \multicolumn{2}{c}{Mean interval width}              & \multicolumn{2}{c}{Median interval width}            \\
\cmidrule(l{3pt}r{3pt}){2-3} \cmidrule(l{3pt}r{3pt}){4-5} \cmidrule(l{3pt}r{3pt}){6-7} \cmidrule(l{3pt}r{3pt}){8-9} \cmidrule(l{3pt}r{3pt}){10-11}
     & \multicolumn{1}{c}{MACP} & \multicolumn{1}{c}{AcMCP} & \multicolumn{1}{c}{MACP} & \multicolumn{1}{c}{AcMCP} & \multicolumn{1}{c}{MACP} & \multicolumn{1}{c}{AcMCP} & \multicolumn{1}{c}{MACP} & \multicolumn{1}{c}{AcMCP} & \multicolumn{1}{c}{MACP} & \multicolumn{1}{c}{AcMCP} \\
\midrule
     & \multicolumn{10}{c}{AR(2) simulation data}                                                                                                                                                                                                                                       \\[0.3em]
h=1  & 90.00 & 90.00 & 89.20 & 89.40 & 90.80 & 90.80 & 3.83 & 3.55 & 3.63 & 3.54 \\
h=2  & 89.99 & 89.94 & 89.20 & 89.00 & 91.00 & 91.00 & 5.17 & 4.68 & 4.91 & 4.66 \\
h=3  & 89.99 & 90.02 & 88.80 & 88.80 & 91.00 & 91.20 & 5.32 & 4.86 & 5.23 & 4.84 \\
\cmidrule(l{3pt}r{3pt}){2-11}
     & \multicolumn{10}{c}{Nonlinear simulation data}                                                                                                                                                                                                                                   \\[0.3em]
h=1  & 90.00 & 90.00 & 87.00 & 86.00 & 93.00 & 93.00 & 0.35 & 0.36 & 0.33 & 0.36 \\
h=2  & 89.98 & 90.18 & 87.00 & 87.00 & 94.00 & 93.00 & 0.41 & 0.37 & 0.34 & 0.36 \\
h=3  & 89.96 & 90.06 & 87.00 & 87.00 & 94.00 & 94.00 & 0.38 & 0.38 & 0.36 & 0.38 \\
\cmidrule(l{3pt}r{3pt}){2-11}
     & \multicolumn{10}{c}{Electricity demand data}                                                                                                                                                                                                                                     \\[0.3em]
h=1  & 90.31 & 89.15 & 86.00 & 85.00 & 92.00 & 92.00 & 26.18 & 24.44 & 22.46 & 24.24 \\
h=2  & 90.62 & 90.23 & 85.00 & 86.00 & 94.00 & 94.00 & 37.87 & 30.72 & 30.92 & 28.87 \\
h=3  & 90.55 & 90.94 & 85.00 & 86.00 & 93.00 & 93.00 & 42.03 & 34.35 & 30.93 & 33.57 \\
h=4  & 90.87 & 90.87 & 86.00 & 88.00 & 96.00 & 94.00 & 43.24 & 40.74 & 33.79 & 39.48 \\
h=5  & 91.20 & 94.80 & 87.00 & 91.00 & 96.00 & 99.00 & 64.01 & 55.62 & 55.19 & 45.43 \\
h=6  & 90.73 & 90.73 & 84.00 & 88.00 & 94.00 & 97.00 & 45.89 & 46.65 & 39.39 & 41.56 \\
h=7  & 90.24 & 89.84 & 85.00 & 87.00 & 93.00 & 95.00 & 50.81 & 47.45 & 35.81 & 47.98 \\
\cmidrule(l{3pt}r{3pt}){2-11}
     & \multicolumn{10}{c}{Eating out expenditure data}                                                                                                                                                                                                                                 \\[0.3em]
h=1  & 92.16 & 88.89 & 90.00 & 83.33 & 96.67 & 93.33 & 67.49 & 65.51 & 64.89 & 63.08 \\
h=2  & 92.05 & 90.07 & 85.00 & 83.33 & 96.67 & 96.67 & 90.12 & 91.50 & 81.98 & 89.10 \\
h=3  & 91.28 & 89.26 & 86.67 & 83.33 & 96.67 & 95.00 & 105.01 & 112.42 & 96.03 & 107.83 \\
h=4  & 91.84 & 89.12 & 86.67 & 81.67 & 98.33 & 98.33 & 132.90 & 116.80 & 120.67 & 111.16 \\
h=5  & 92.41 & 91.03 & 85.00 & 81.67 & 98.33 & 96.67 & 133.90 & 126.12 & 115.52 & 120.97 \\
h=6  & 91.61 & 90.91 & 85.00 & 85.00 & 98.33 & 96.67 & 135.21 & 178.17 & 125.62 & 178.60 \\
h=7  & 88.65 & 86.52 & 78.33 & 73.33 & 98.33 & 98.33 & 140.76 & 186.58 & 140.97 & 198.27 \\
h=8  & 87.77 & 88.49 & 71.67 & 83.33 & 100.00 & 98.33 & 148.63 & 192.07 & 148.90 & 192.59 \\
h=9  & 86.13 & 89.78 & 68.33 & 86.67 & 100.00 & 100.00 & 156.46 & 194.40 & 161.81 & 168.96 \\
h=10 & 88.15 & 89.63 & 75.00 & 85.00 & 100.00 & 93.33 & 166.56 & 195.83 & 170.70 & 193.68 \\
h=11 & 85.71 & 88.72 & 70.00 & 85.00 & 100.00 & 96.67 & 163.92 & 233.80 & 158.95 & 220.92 \\
h=12 & 85.50 & 91.60 & 68.33 & 83.33 & 100.00 & 98.33 & 168.05 & 336.02 & 181.64 & 297.48 \\
\bottomrule
\end{tabular}}

}

\end{table}%

\section{Conclusion and discussion}\label{conclusion-and-discussion}

We propose MPID and AcMCP to construct prediction sets for multi-step
time series forecasting. Both methods extend conformal PID control
\citep{angelopoulos2024} to accommodate delayed feedback inherent in
this setting, while AcMCP additionally incorporates dependence among
forecast errors across horizons into its scorecaster. Decompositions of
optimal forecast errors, together with explicit bounds on Taylor
remainders, provide a structural rationale for this design and clarify
the conditions supporting the underlying linear approximations. For both
methods, we establish finite-sample bounds and horizon-specific long-run
guarantees for empirical coverage under the stated calibration
conditions, without requiring exchangeability, stationarity, or
assumptions on how the data distribution evolves over time. These
guarantees do not depend on the accuracy of the linear approximations.
The coverage bounds distinguish the contribution of calibration feedback
from that of forecast delay, whose effect vanishes asymptotically for
each fixed horizon. Across the experiments considered, MPID and AcMCP
adapt their interval widths to changing time series dynamics, generally
maintaining more stable rolling coverage around the target than the
benchmarks. AcMCP often produces narrower intervals than MPID with
similar coverage, indicating the empirical value of incorporating
forecast error dependence into conformal calibration. This advantage is
not uniform, however: when the step size of MACP is tuned post hoc on
the test set, MACP produces intervals comparable to those of AcMCP and,
at longer horizons in the real data sets, sometimes narrower ones with
similar coverage.

Our analysis of forecast errors is restricted to an ex-post forecasting
setting, in which future values of exogenous predictors are assumed to
be known. In this setting, interval width reflects both \emph{aleatoric
uncertainty}, arising from intrinsic randomness in the DGP, and
\emph{epistemic uncertainty}, stemming from a lack of knowledge about
the underlying DGP \citep{sale2025}. In many practical forecasting
scenarios, however, predictors must themselves be forecast. Uncertainty
in future predictor values and in the models used to forecast them can
then propagate into response forecasts, motivating explicit treatment of
predictor uncertainty in conformal inference.

Although the forecast error analysis assumes known future predictors,
the coverage guarantees remain applicable when these predictors are
forecast, provided that the assumptions of Corollary~\ref{cor-cov_acmcp}
and Corollary~\ref{cor-cov-two-sided} hold. Appendix
\ref{sec-predictor-uncertainty} compares AcMCP using known and forecast
future predictors in a simulation. Both settings achieve near-nominal
empirical coverage and small rolling coverage errors, but forecasting
the predictors produces wider intervals, with larger increases at higher
levels of predictor uncertainty and longer forecast horizons.

These considerations suggest several directions for future research. One
is to improve interval efficiency while preserving coverage guarantees.
Our methods prioritize efficient online implementation through simple
updates. A complementary approach is Bellman conformal inference
\citep{yang2024ts}, which uses a look-ahead optimization to balance
interval length against a miscoverage penalty, but issues only a
one-step-ahead interval. Extending this optimization principle to
multi-step interval construction while retaining coverage guarantees is
a promising direction. Automated learning-rate selection could further
reduce manual tuning and improve adaptation to changing forecasting
conditions. Other avenues include establishing conditional coverage
guarantees under explicit restrictions on distributional change and
developing a theoretical understanding of how uncertainty in future
predictors and their forecasting models affects interval efficiency and
local calibration.

% \newpage
\appendix
% \pagenumbering{arabic}% resets `page` counter to 1
\setcounter{section}{0}
\renewcommand{\thesection}{Appendix \Alph{section}}
\renewcommand{\thesubsection}{\Alph{section}.\arabic{subsection}}

\section*{Acknowledgments}\label{acknowledgments}
\addcontentsline{toc}{section}{Acknowledgments}

This work was supported by the National Natural Science Foundation of
China (No.~72601399), the Presidential Foundation of the Academy of
Mathematics and Systems Science, Chinese Academy of Sciences, China
(No.~E555930101), and the Australian Research Council Industrial
Transformation Training Centre in Optimisation Technologies, Integrated
Methodologies, and Applications (OPTIMA) (No.~IC200100009).

\section*{Disclosure statement}\label{disclosure-statement}
\addcontentsline{toc}{section}{Disclosure statement}

The authors report there are no competing interests to declare.

\section*{Supplementary materials}\label{supplementary-materials}
\addcontentsline{toc}{section}{Supplementary materials}

The Appendix provides proofs of all propositions and corollaries,
additional multi-step conformal methods, analyses of approximation
errors and local coverage, supplementary two-sided coverage results, and
a simulation examining the effect of uncertainty in future predictors.

\section*{Declaration of generative AI
use}\label{declaration-of-generative-ai-use}
\addcontentsline{toc}{section}{Declaration of generative AI use}

During the preparation of this work the authors used OpenAI Codex (GPT-6
Astra, OpenAI) and Claude (Opus 5.5, Anthropic) to correct typographical
errors and improve the language. After using these tools, the authors
reviewed and edited the content as needed and take full responsibility
for the content of the publication.

\section*{References}\label{references}
\addcontentsline{toc}{section}{References}

\begingroup

\renewcommand{\bibsection}{}
\bibliography{references.bib}

\endgroup

\newpage
\appendix
\pagenumbering{arabic}% resets `page` counter to 1

\setcounter{section}{0}
\setcounter{subsection}{0}
\setcounter{figure}{0}
\setcounter{table}{0}

\renewcommand{\thesection}{\Alph{section}}
\renewcommand{\thesubsection}{A\arabic{subsection}}
\renewcommand{\thefigure}{A\arabic{figure}}
\renewcommand{\thetable}{A\arabic{table}}

\section*{Appendix}\label{appendix}
\addcontentsline{toc}{section}{Appendix}

\subsection{Multi-step extensions of existing conformal
methods}\label{sec-extensions}

\subsubsection{Online multi-step split conformal
prediction}\label{online-multi-step-split-conformal-prediction}

Split conformal prediction \citep[SCP,][]{papadopoulos2002, vovk2005} is
a holdout method for building prediction sets using a model pre-trained
on a training set. A key advantage of SCP is its ability to guarantee
coverage by assuming data exchangeability. In a regression setting, SCP
randomly separates the available \(n\) data points,
\(Z_i = (X_i, Y_i) \in \mathbb{R}^d \times \mathbb{R}\),
\(i=1, \dots, n\), into a proper training set of size \(n_t\) and a
calibration set of size \(n_c\). Given a regression model
\(\hat{\mu}: \mathbb{R}^d \rightarrow \mathbb{R}\) fitted on the
training set and a score function \(\mathcal{S}\), nonconformity scores
\(s_i = \mathcal{S}\left(X_i, Y_i\right)\) are computed on the
calibration set to measure the nonconformity between the calibration
response values and the predicted values obtained from the fitted model
\(\hat{\mu}\). Then SCP computes the prediction set for the test
response \(Y_{n+1}\) using \[
\hat{\mathcal{C}}_{n+1}^{\alpha}\left(X_{n+1}\right) = \left\{y\in\mathbb{R}: \mathcal{S}\left(X_{n+1}, y\right) \leq Q_{1-\alpha}\left(\sum_{i \in \mathcal{D}_{\text{cal}}}\frac{1}{n_c+1}\cdot\delta_{s_{i}}+\frac{1}{n_c+1}\cdot\delta_{+\infty}\right)\right\},
\] where \(Q_\tau(\cdot)\) denotes the \(\tau\)-quantile of its
argument, and \(\delta_a\) denotes the point mass at \(a\). Time series
data are inherently nonexchangeable due to their temporal dependence and
autocorrelation. Therefore, directly applying SCP to time series data
would violate the method's exchangeability assumption and compromise its
coverage guarantee.

Here we introduce online \textbf{multi-step split conformal prediction}
(MSCP) as a generalization of SCP \citep{papadopoulos2002, vovk2005} to
recursively update all \(h\)-step-ahead prediction intervals over time.
MSCP applies conformal inference in an online fashion, updating
prediction intervals as new data points are received. Specifically, for
each \(h \in [H]\), the one-sided prediction set is
\begin{equation}\phantomsection\label{eq-mscp}{
\hat{\mathcal{C}}_{t+h|t}^{\alpha,\mathrm{tail}} = \Bigg\{y\in\mathbb{R}: s_{t+h|t}^{y} \leq Q_{1-\alpha}\Bigg(\sum_{i=t-t_c+1}^{t}\frac{1}{t_c+1}\cdot\delta_{s_{i|i-h}}+\frac{1}{t_c+1}\cdot\delta_{+\infty}\Bigg)\Bigg\},
}\end{equation} where \(s_{t+h|t}^{y}\) is the relevant signed
nonconformity score defined in Section~\ref{sec-setup}. Applying
Equation \eqref{eq-mscp} to both tails and combining the resulting
thresholds as in Equation \eqref{eq-two-sided-pi} yields the MSCP
interval.

\subsubsection{Online multi-step weighted conformal
prediction}\label{online-multi-step-weighted-conformal-prediction}

\citet{barber2023} propose a nonexchangeable conformal prediction
procedure (NexCP) that generalizes SCP to certain nonexchangeable
settings by assigning higher weights to observations believed to share
the same distribution as the test data. NexCP assumes these weights are
fixed and data-independent. Under exchangeability, NexCP retains the
same coverage guarantees as SCP; when exchangeability is violated, the
coverage gap is governed by the total variation distance between swapped
nonconformity score vectors, which can be substantial in time series
settings.

The online \textbf{multi-step weighted conformal prediction} (MWCP)
method we propose here adapts the NexCP method to the online setting for
time series forecasting. MWCP uses weighted quantile estimates for
constructing prediction intervals, contrasting with the MSCP definitions
where all nonconformity scores for calibration are implicitly assigned
equal weight.

For the subsequent empirical study, we choose fixed weights
\(w_i = b^{t+1-i}\), \(b \in (0, 1)\) and \(i=t-t_c+1, \dots, t\), for
nonconformity scores on the corresponding calibration set. In this
setting, weights decay exponentially as the nonconformity scores get
older, akin to the rationale behind the simple exponential smoothing
method in time series forecasting \citep{hyndman2021}. Then, for each
\(h \in [H]\), MWCP updates the \(h\)-step-ahead one-sided prediction
set: \[
\hat{\mathcal{C}}_{t+h|t}^{\alpha,\mathrm{tail}} = \Bigg\{y\in\mathbb{R}: s_{t+h|t}^{y} \leq Q_{1-\alpha}\Bigg(\sum_{i=t-t_c+1}^{t}\tilde{w}_i\cdot\delta_{s_{i|i-h}}+\tilde{w}_{t+1}\cdot\delta_{+\infty}\Bigg)\Bigg\},
\] where \(\tilde{w}_i\) and \(\tilde{w}_{t+1}\) are normalized weights
given by \[
\tilde{w}_i = \frac{w_i}{\sum_{i=t-t_c+1}^{t}w_i+1}, \text{ for } i \in \{t-t_c+1, \dots, t\} \quad \text{and} \quad \tilde{w}_{t+1} =  \frac{1}{\sum_{i=t-t_c+1}^{t}w_i+1}.
\] As with MSCP, the two one-sided MWCP sets are combined according to
Equation \eqref{eq-two-sided-pi}.

\subsubsection{Multi-step adaptive conformal
prediction}\label{multi-step-adaptive-conformal-prediction}

Next we extend the adaptive conformal inference (ACI) method proposed by
\citet{gibbs2021} to address multi-step time series forecasting,
introducing the \textbf{multi-step adaptive conformal prediction} (MACP)
method. For a generic tail with fixed target miscoverage level
\(\alpha\), the method adapts the nominal level \(\alpha_{t+h|t}\) used
for quantile estimation to past tail miscoverage indicators.
Specifically, for each \(h \in [H]\), we update \(\alpha_{t+h|t}\)
according to \begin{equation}\phantomsection\label{eq-macp}{
\alpha_{t+h|t} := \alpha_{t+h-1|t-1} + \gamma(\alpha - \mathrm{err}_{t|t-h}).
}\end{equation} The corresponding one-sided set is computed from
Equation \eqref{eq-mscp} with \(\alpha\) replaced by \(\alpha_{t+h|t}\).
For levels at or below zero, the one-sided set is \(\mathbb R\); for
levels at or above one, it is empty. Here, \(\gamma > 0\) is a fixed
step size, and \(\alpha_{2h|h}\) is the initial estimate, typically set
to \(\alpha\). The upper- and lower-tail recursions are run separately
and then combined through Equation \eqref{eq-two-sided-pi}.

Equation \eqref{eq-macp} uses the tail miscoverage indicator observed at
time \(t\) to update the forecast for time \(t+h\). The nominal tail
miscoverage level increases when the observation satisfies the relevant
upper or lower bound (\(s_{t|t-h}\leq q_{t|t-h}\)) and decreases when it
violates that bound (\(s_{t|t-h}>q_{t|t-h}\)). For a fixed calibration
sample, decreasing this level widens the one-sided prediction set.
Equation \eqref{eq-macp} reduces to the ACI update when \(h=1\).

We do not use the update
\(\alpha_{t+1|t-h+1} := \alpha_{t|t-h} + \gamma(\alpha - \mathrm{err}_{t|t-h})\),
because it requires \(y_t\), which is unavailable at the forecast origin
\(t-h+1\) when \(h>1\).

Selecting the parameter \(\gamma\) is pivotal yet challenging.
\citet{gibbs2021} relate the choice of \(\gamma\) to the rate of
distributional change. Several strategies have been proposed to avoid
the necessity of selecting \(\gamma\). For example, \citet{zaffran2022}
use an adaptive aggregation of multiple ACIs with a set of candidate
values for \(\gamma\), determining weights based on their historical
performance. \citet{bastani2022} propose a multivalid prediction
algorithm in which the prediction set is established by selecting a
threshold from a sequence of candidate thresholds. Both methods rely on
update schemes that place substantial weight on older historical data,
which may impede rapid adaptation to abrupt changes \citep{gibbs2024}.
Therefore, \citet{gibbs2024} propose an alternative expert selection
scheme that adaptively tunes the step-size parameter over time and
places greater emphasis on more recent observations by construction,
enabling faster responses to sudden environmental shifts.

The theoretical coverage properties of ACI suggest that a larger value
for \(\gamma\) generally results in less deviation from the target
coverage. As there is no restriction on \(\alpha_{t+h|t}\), it can drift
below \(0\) or above \(1\); a larger \(\gamma\) may therefore lead to
frequent empty or unbounded one-sided sets as the procedure responds to
whether the relevant bound is violated.

\subsection{\texorpdfstring{Proof of
Proposition~\ref{prp-ar}}{Proof of Proposition~}}\label{sec-proof_ar}

\begin{proof}
Fix a forecast origin \(t\). For \(h=1\), all lagged responses and
required predictors are \(\mathcal F_t\)-measurable. Since
\(\mathbb E(\varepsilon_{t+1}|\mathcal F_t)=0\), we have \[
\hat y_{t+1|t}=f_{t+1}(y_{(t+1-d):t},\bm x_{(t+1-k):(t+1)})
\] and hence \(e_{t+1|t}=\varepsilon_{t+1}\), with
\(\widetilde R_{t,1}=0\).

For \(h\geq2\), write
\(\bm\Delta_{t,h}=\bm z_{t,h}-\bm m_{t,h}=(e_{t+h-1|t},\dots,e_{t+h-p_h|t})^{\prime}\).
Applying a first-order Taylor expansion and retaining the remainder
gives \begin{align*}
y_{t+h}
&=F_{t,h}(\bm z_{t,h})+\varepsilon_{t+h}\\
&=F_{t,h}(\bm m_{t,h})
+\sum_{i=1}^{p_h}\partial_iF_{t,h}(\bm m_{t,h})e_{t+h-i|t}
+R_{t,h}+\varepsilon_{t+h},
\end{align*} where \(R_{t,h}\) is the Taylor remainder beyond first
order.

The expansion point and derivative coefficients are
\(\mathcal F_t\)-measurable. Since
\(\mathbb E(e_{t+h-i|t}|\mathcal F_t)=0\) for \(1\leq i\leq p_h\) and,
by the tower property, \(\mathbb E(\varepsilon_{t+h}|\mathcal F_t)=0\),
taking conditional expectations under the stated integrability
assumptions gives \[
\hat y_{t+h|t}=F_{t,h}(\bm m_{t,h})+\mathbb E(R_{t,h}|\mathcal F_t).
\] Subtracting this identity from the expansion yields \[
e_{t+h|t}=\varepsilon_{t+h}+\sum_{i=1}^{p_h}\phi_{t,h,i}e_{t+h-i|t}+\widetilde R_{t,h},
\] where \(\widetilde R_{t,h}=R_{t,h}-\mathbb E(R_{t,h}|\mathcal F_t)\)
and \(\mathbb E(\widetilde R_{t,h}|\mathcal F_t)=0\). Together with the
case \(h=1\), interpreting an empty sum as zero, this proves Equation
\eqref{eq-ar}. If \(f_{t+h}\) is linear in its lagged responses,
allowing an intercept, then \(R_{t,h}=\widetilde R_{t,h}=0\).

To establish Equation \eqref{eq-remainder-bound}, use the integral
representation \[
R_{t,h}=\int_0^1(1-u)\bm\Delta_{t,h}^{\prime}
\nabla^2 F_{t,h}(\bm m_{t,h}+u\bm\Delta_{t,h})\bm\Delta_{t,h}\,du,
\] where \(\nabla^2F_{t,h}\) is the Hessian with respect to the
unobserved response coordinates. The Hessian bound implies
\(|R_{t,h}|\leq(M_{t,h}/2)\|\bm\Delta_{t,h}\|_2^2\). Conditional
Jensen's inequality and the triangle inequality give \[
\mathbb E\!\left[|\widetilde R_{t,h}|\,|\,\mathcal F_t\right]
\leq2\mathbb E\!\left[|R_{t,h}|\,|\,\mathcal F_t\right]
\leq M_{t,h}\mathbb E\!\left[\|\bm\Delta_{t,h}\|_2^2\,|\,\mathcal F_t\right].
\] Since \(\|\bm\Delta_{t,h}\|_2^2=\sum_{i=1}^{p_h}e_{t+h-i|t}^2\), this
proves Equation \eqref{eq-remainder-bound}.
\end{proof}

\subsection{\texorpdfstring{Proof of
Proposition~\ref{prp-ma}}{Proof of Proposition~}}\label{sec-proof_ma}

\begin{proof}
Fix a forecast origin \(t\). By Proposition~\ref{prp-ar}, \begin{align*}
e_{t+1|t}&=\varepsilon_{t+1},\\
e_{t+j|t}&=\varepsilon_{t+j}+\sum_{i=1}^{p_j}\phi_{t,j,i}e_{t+j-i|t}+\widetilde R_{t,j},\qquad 2\leq j\leq h.
\end{align*}

Recursively substituting the earlier-horizon relations gives \[
e_{t+h|t}=\varepsilon_{t+h}+\sum_{i=1}^{h-1}\theta_{t,h,i}\varepsilon_{t+h-i}+U_{t,h}.
\] The coefficients \(\theta_{t,h,i}\) are finite sums of products of
the \(\phi\) coefficients and are therefore \(\mathcal F_t\)-measurable.
Collecting the remainder terms yields
\begin{equation}\phantomsection\label{eq-propagated-remainder}{
U_{t,1}=0,\qquad
U_{t,j}=\widetilde R_{t,j}+\sum_{i=1}^{p_j}\phi_{t,j,i}U_{t,j-i},\qquad 2\leq j\leq h.
}\end{equation} This proves Equation \eqref{eq-ma} without discarding
any Taylor remainder.

Under the additional assumptions of Proposition~\ref{prp-ma}, Equation
\eqref{eq-remainder-bound} gives
\(\mathbb E(|\widetilde R_{t,j}|\,|\,\mathcal F_t)\leq MV\) for
\(2\leq j\leq h\). The triangle inequality and
\(\mathcal F_t\)-measurability of the coefficients imply \[
\mathbb E(|U_{t,j}|\,|\,\mathcal F_t)
\leq MV+\sum_{i=1}^{p_j}|\phi_{t,j,i}|
\mathbb E(|U_{t,j-i}|\,|\,\mathcal F_t).
\] Let \(G_1=0\) and \(G_j=\sum_{k=0}^{j-2}L^k\) for \(j\geq2\). The
bound \(\mathbb E(|U_{t,1}|\,|\,\mathcal F_t)\leq MVG_1\) holds since
\(U_{t,1}=0\). If it holds for all horizons below \(j\), then
\begin{align*}
\mathbb E(|U_{t,j}|\,|\,\mathcal F_t)
&\leq MV\left(1+\sum_{i=1}^{p_j}|\phi_{t,j,i}|G_{j-i}\right)\\
&\leq MV(1+LG_{j-1})=MVG_j,
\end{align*} where \(G_{j-i}\leq G_{j-1}\) because \(L\geq0\). Strong
induction therefore establishes the bound for all \(1\leq j\leq h\), and
taking \(j=h\) proves Equation \eqref{eq-propagated-bound}.
\end{proof}

\subsection{\texorpdfstring{Approximation error analysis for
Propositions \ref{prp-ar} and
\ref{prp-ma}}{Approximation error analysis for Propositions  and }}\label{sec-approximation-errors}

Omitting the remainders in Propositions \ref{prp-ar} and \ref{prp-ma}
yields linear approximations with errors \(\widetilde R_{t,h}\) and
\(U_{t,h}\). Equation \eqref{eq-remainder-bound} controls the first
through curvature and shorter-horizon conditional mean squared errors;
Equation \eqref{eq-propagated-bound} additionally accounts for
propagation through \(G_h\), defined in Appendix \ref{sec-proof_ma}. For
fixed \(L\geq0\), this factor is bounded by \((1-L)^{-1}\) when \(L<1\),
equals \(h-1\) when \(L=1\), and grows geometrically with \(h\) when
\(L>1\).

For the small-perturbation analysis, fix \(t,h,d\) with \(h\geq2\) and
consider a family of processes indexed by \(\delta>0\), each satisfying
the assumptions for Equation \eqref{eq-ar} in Proposition~\ref{prp-ar}.
Suppose that the following bounds hold almost surely for finite
deterministic constants \(M,K,\sigma\geq0\) independent of \(\delta\):

\begin{enumerate}
\def\labelenumi{\arabic{enumi}.}
\tightlist
\item
  The Hessian bounds in Proposition~\ref{prp-ar} hold with
  \(M_{t,j}\leq M\) for \(2\leq j\leq h\).
\item
  The gradients along the expansion segments satisfy \[
  \sup_{u\in[0,1]}\left\|\nabla F_{t,j}\bigl(\bm m_{t,j}+u(\bm z_{t,j}-\bm m_{t,j})\bigr)\right\|_2\leq K,
  \qquad 2\leq j\leq h.
  \]
\item
  The innovation conditional variances satisfy
  \(\mathbb E(\varepsilon_{t+j}^2|\mathcal F_{t+j-1})\leq\sigma^2\delta^2\)
  for \(1\leq j\leq h\).
\end{enumerate}

\noindent Dependence on \(\delta\) is suppressed in the notation.

By Cauchy--Schwarz, \(\sum_{i=1}^{p_j}|\phi_{t,j,i}|\leq\sqrt d K\), so
\(L=\sqrt d K\) is admissible.

Write \(V_j=\mathbb E(e_{t+j|t}^2|\mathcal F_t)\) for \(1\leq j\leq h\).
Conditional mean optimality gives, for \(j\geq2\), \[
V_j\leq\mathbb E\!\left((y_{t+j}-F_{t,j}(\bm m_{t,j}))^2\,|\,\mathcal F_t\right).
\] Since \(F_{t,j}(\bm z_{t,j})-F_{t,j}(\bm m_{t,j})\) is
\(\mathcal F_{t+j-1}\)-measurable, its cross term with
\(\varepsilon_{t+j}\) has conditional expectation zero. The gradient
bound gives
\(|F_{t,j}(\bm z_{t,j})-F_{t,j}(\bm m_{t,j})|\leq K\|\bm\Delta_{t,j}\|_2\),
using the increment defined in Appendix \ref{sec-proof_ar}. Together
with the innovation variance bound, this yields \[
V_1\leq\sigma^2\delta^2,\qquad
V_j\leq\sigma^2\delta^2+K^2\sum_{i=1}^{p_j}V_{j-i},\qquad 2\leq j\leq h.
\] Induction gives \(V_j\leq A_j\delta^2\), where \(A_1=\sigma^2\) and
\(A_j=\sigma^2+K^2\sum_{i=1}^{p_j}A_{j-i}\). Taking
\(C=\max_{2\leq j\leq h}\sum_{i=1}^{p_j}A_{j-i}\), independent of
\(\delta\), and substituting \(V=C\delta^2\) into the remainder bounds
yields, almost surely, \[
\mathbb E(|\widetilde R_{t,j}|\,|\,\mathcal F_t)\leq MC\delta^2,\qquad 2\leq j\leq h,
\qquad
\mathbb E(|U_{t,h}|\,|\,\mathcal F_t)\leq MCG_h\delta^2.
\] Taking expectations shows that, under the stated conditions, both
remainders have mean absolute size \(O(\delta^2)\) and are negligible
relative to \(\delta\). Specifically,
\(\mathbb E|\widetilde R_{t,j}/\delta|\to0\) for \(2\leq j\leq h\) and
\(\mathbb E|U_{t,h}/\delta|\to0\) as \(\delta\downarrow0\).

These results quantify linear approximation errors as the innovation
scale decreases at fixed forecast horizons, rather than as the
training-sample size increases. At a given perturbation scale, large
curvature or propagation factors may make the bounds uninformative.

\subsection{\texorpdfstring{Proof of
Proposition~\ref{prp-cov_rt}}{Proof of Proposition~}}\label{sec-proof_cov_rt}

\begin{proof}
Let \(E_T=\sum_{i=h+1}^{T}(\mathrm{err}_{i|i-h}-\alpha)\). The
inequality in Equation \eqref{eq-cov_rt} is equivalent to
\(|E_T|\leq cg(T-h)+h\). We first prove the upper bound by strong
induction, with the lower bound following analogously.

For \(T=h+1,\dots,2h\), \(|E_T|\leq T-h\leq h\leq cg(T-h)+h\), since
\(g\) is nonnegative and \(0\leq\mathrm{err}_{i|i-h}\leq1\). Thus both
bounds hold over the initial range.

Now let \(T\geq2h+1\) and assume the upper bound holds at all earlier
indices. The threshold evaluated at time \(T\) is \[
q_{T|T-h}=r_{T-h}(E_{T-h}).
\] We partition the argument into two cases. If
\(E_{T-h}\geq cg(T-2h)\), Equation \eqref{eq-saturation_h} gives
\(q_{T|T-h}\geq b\), so \(\mathrm{err}_{T|T-h}=0\). By the induction
hypothesis and monotonicity of \(g\), \[
E_T=E_{T-1}-\alpha\leq cg(T-1-h)+h\leq cg(T-h)+h.
\] Otherwise, using \(\mathrm{err}_{i|i-h}-\alpha\leq1-\alpha\) over the
last \(h\) observations gives \[
E_T=E_{T-h}+\sum_{i=T-h+1}^{T}(\mathrm{err}_{i|i-h}-\alpha)
<cg(T-2h)+h(1-\alpha)\leq cg(T-h)+h.
\] This completes the induction for the upper bound. For the lower
bound, the same induction applies: if \(E_{T-h}\leq-cg(T-2h)\),
saturation and \(s_{T|T-h}>-b\) imply \(\mathrm{err}_{T|T-h}=1\), so \[
E_T=E_{T-1}+1-\alpha\geq-cg(T-1-h)-h\geq-cg(T-h)-h.
\] Otherwise, \(\mathrm{err}_{i|i-h}-\alpha\geq-\alpha\) yields \[
E_T>-cg(T-2h)-h\alpha\geq-cg(T-h)-h.
\] The argument also applies when \(b=\infty\) because every score is
finite. Dividing by \(T-h\) proves Equation \eqref{eq-cov_rt}.

Regarding the long-run conclusion, \(g(T-h)/(T-h)\to0\) by sublinearity,
and \(h/(T-h)\to0\) for fixed \(h\). Hence the empirical tail
miscoverage rate converges to \(\alpha\).
\end{proof}

\subsection{\texorpdfstring{Proof of
Corollary~\ref{cor-cov_qt}}{Proof of Corollary~}}\label{sec-proof_cov_qt}

\begin{proof}
Unraveling the recursion gives \[
q_{t+h|t}=q_{2h|h}+\eta\sum_{i=h+1}^{t}(\mathrm{err}_{i|i-h}-\alpha).
\] Thus, take \(r_t(x)=q_{2h|h}+\eta x\), \(g\equiv1\), and
\(c=(b+|q_{2h|h}|)/\eta\). If \(x\geq c\), then \(r_t(x)\geq b\), and if
\(x\leq-c\), then \(r_t(x)\leq-b\), so Equation \eqref{eq-saturation_h}
holds. Applying Proposition~\ref{prp-cov_rt} gives the stated bound and
long-run calibration.
\end{proof}

\subsection{\texorpdfstring{Proof of
Corollary~\ref{cor-cov_acmcp}}{Proof of Corollary~}}\label{sec-proof_cov_acmcp}

\begin{proof}
Let \(s_{t+h|t}^{*}=s_{t+h|t}-\hat q_{t+h|t}\) and
\(q_{t+h|t}^{*}=q_{t+h|t}-\hat q_{t+h|t}\). Since the scores and
baseline values are finite, these transformations are well-defined even
when the thresholds are infinite, and \(s_{t+h|t}^{*}\in(-b,b)\) by
assumption. Subtracting the same finite baseline preserves the tail
miscoverage indicator:
\(\mathrm{err}_{t+h|t}=\mathbbm{1}\{s_{t+h|t}^{*}>q_{t+h|t}^{*}\}\).
Equation \eqref{eq-acmcp_1} therefore becomes \[
q_{t+h|t}^{*}=r_t\left(\sum_{i=h+1}^{t}\bigl(\mathbbm{1}\{s_{i|i-h}^{*}>q_{i|i-h}^{*}\}-\alpha\bigr)\right).
\] The transformed recursion satisfies the assumptions of
Proposition~\ref{prp-cov_rt} with the same \(b\), \(c\), and \(g\).
Applying that proposition gives the stated finite-sample bound and
long-run calibration for the unchanged indicators.
\end{proof}

\subsection{\texorpdfstring{Proof of
Corollary~\ref{cor-cov-two-sided}}{Proof of Corollary~}}\label{sec-proof_cov_two_sided}

\begin{proof}
Common saturation constants can be chosen by taking the maximum of the
two constants and the pointwise maximum of the two admissible functions.
Apply Corollary~\ref{cor-cov_acmcp} separately to the lower- and
upper-tail recursions, replacing its generic target \(\alpha\) by
\(\alpha^{-}\) and \(\alpha^{+}\), respectively. With \(n=T-h\), this
gives \[
\left|\frac{1}{n}\sum_{t=h+1}^{T}\mathrm{err}_{t|t-h}^{a}-\alpha^{a}\right|\leq \frac{c\cdot g(n)+h}{n},
\qquad a\in\{-,+\}.
\] The interval fails to cover \(y_t\) if and only if at least one tail
constraint fails, including when the interval is empty. Thus, \[
\mathbbm{1}\{y_t\notin\hat{\mathcal C}_{t|t-h}^{\alpha}\}
=\mathrm{err}_{t|t-h}^{-}+\mathrm{err}_{t|t-h}^{+}-\mathrm{err}_{t|t-h}^{-}\mathrm{err}_{t|t-h}^{+}.
\] Dropping the nonnegative overlap term, summing, and using
\(\alpha^{-}+\alpha^{+}=\alpha\) yields \[
\frac{1}{n}\sum_{t=h+1}^{T}\mathbbm{1}\{y_t\notin\hat{\mathcal C}_{t|t-h}^{\alpha}\}
\leq \alpha+2\frac{c\cdot g(n)+h}{n},
\] which is equivalent to Equation \eqref{eq-two-sided-finite}. Because
\(g\) is sublinear and \(h\) is fixed,
\(\bigl(c\cdot g(n)+h\bigr)/n\to0\), giving the long-run coverage lower
bound.

For the absolute deviation bound, simultaneous violations of both tail
bounds imply \[
q_{t|t-h}^{+}<y_t-\hat y_{t|t-h}< -q_{t|t-h}^{-},
\] so \(\hat y_{t|t-h}-q_{t|t-h}^{-}>\hat y_{t|t-h}+q_{t|t-h}^{+}\), a
crossing of the displayed endpoints. This comparison remains
well-defined for infinite thresholds. Thus the average of the products
of the two tail indicators is at most \(\chi_{T,h}\). The union identity
and the triangle inequality give \[
\begin{aligned}
\left|\frac1n\sum_{t=h+1}^{T}\mathbbm{1}\{y_t\in\hat{\mathcal C}_{t|t-h}^{\alpha}\}-(1-\alpha)\right|
&\leq\sum_{a\in\{-,+\}}\left|\frac1n\sum_{t=h+1}^{T}\mathrm{err}_{t|t-h}^{a}-\alpha^a\right|
+\frac1n\sum_{t=h+1}^{T}\mathrm{err}_{t|t-h}^{-}\mathrm{err}_{t|t-h}^{+}\\
&\leq2\frac{c\cdot g(n)+h}{n}+\chi_{T,h}.
\end{aligned}
\] This proves Equation \eqref{eq-two-sided-enhanced} and exact long-run
calibration when \(\chi_{T,h}\to0\).

For the iteration in Proposition~\ref{prp-cov_rt}, the same proof
applies with zero baselines.
\end{proof}

\subsection{Two-sided coverage for constant-step quantile
tracking}\label{sec-two-sided-constant-step}

\begin{corollary}[]\protect\hypertarget{cor-cov-two-sided-qt}{}\label{cor-cov-two-sided-qt}

Fix \(h\in[H]\) and targets \(\alpha^{-},\alpha^{+}>0\) with
\(\alpha^{-}+\alpha^{+}=\alpha\in(0,1)\). Apply the iteration in
Corollary~\ref{cor-cov_qt} separately to the two signed tail scores and
construct the interval in Equation \eqref{eq-two-sided-pi}. Suppose each
tail \(a\in\{-,+\}\) satisfies the assumptions of
Corollary~\ref{cor-cov_qt} with finite score bound \(b^a\), step size
\(\eta^a>0\), and finite initial threshold \(q^a_{2h|h}\). Then, for
every \(T\geq h+1\),
\begin{equation}\phantomsection\label{eq-two-sided-constant-step}{
\frac{1}{T-h}\sum_{t=h+1}^{T}\mathbbm{1}\{y_t\in\hat{\mathcal C}_{t|t-h}^{\alpha}\}
\geq 1-\alpha-\sum_{a\in\{-,+\}}\frac{b^a+|q^a_{2h|h}|+\eta^a h}{\eta^a(T-h)}.
}\end{equation} The absolute deviation of the empirical coverage from
\(1-\alpha\) is bounded by the sum on the right plus the
endpoint-crossing frequency \(\chi_{T,h}\) defined in
Corollary~\ref{cor-cov-two-sided}. Consequently, the limit inferior of
the empirical coverage is at least \(1-\alpha\), and the coverage
converges to \(1-\alpha\) if \(\chi_{T,h}\to0\).

\end{corollary}

\begin{proof}
With \(n=T-h\), Corollary~\ref{cor-cov_qt} gives \[
\left|\frac1n\sum_{t=h+1}^{T}\mathrm{err}_{t|t-h}^{a}-\alpha^a\right|
\leq\frac{b^a+|q^a_{2h|h}|+\eta^a h}{\eta^a n},\qquad a\in\{-,+\}.
\] Applying the union identity and endpoint-crossing argument in
Appendix \ref{sec-proof_cov_two_sided} gives Equation
\eqref{eq-two-sided-constant-step} and the absolute deviation bound. For
fixed \(h\), the two tail bounds tend to zero, giving the stated
long-run conclusions.
\end{proof}

\subsection{Local coverage bounds}\label{sec-local-coverage}

Fix \(h\in[H]\) and a window \(\{a+1,\ldots,a+m\}\), where \(a\geq h\)
and \(m\geq1\) are integers. Under the assumptions of
Corollary~\ref{cor-cov_acmcp}, set \(D_h=0\) and \(D_u=c\cdot g(u-h)+h\)
for \(u\geq h+1\). Differencing the cumulative deviations at \(a+m\) and
\(a\) and applying the triangle inequality yields
\begin{equation}\phantomsection\label{eq-window-tail}{
\left|\frac1m\sum_{t=a+1}^{a+m}(\mathrm{err}_{t|t-h}-\alpha)\right|
\leq\frac{D_{a+m}+D_a}{m}.
}\end{equation} Under the assumptions of
Corollary~\ref{cor-cov-two-sided}, with common \(c\) and \(g\) for both
tails and \(\alpha=\alpha^{-}+\alpha^{+}\), the union bound gives
\begin{equation}\phantomsection\label{eq-window-coverage}{
\frac1m\sum_{t=a+1}^{a+m}\mathbbm{1}\{y_t\in\hat{\mathcal C}_{t|t-h}^{\alpha}\}
\geq1-\alpha-2\frac{D_{a+m}+D_a}{m}.
}\end{equation} If \((D_{a+m}+D_a)/m\to0\) along a sequence of windows,
empirical window coverage has limit inferior at least \(1-\alpha\) and
converges to \(1-\alpha\) if the within-window endpoint-crossing
frequency also vanishes, by Appendix \ref{sec-proof_cov_two_sided}.
These bounds may be conservative and uninformative for fixed-length
windows.

\subsection{Sensitivity to uncertainty in future
predictors}\label{sec-predictor-uncertainty}

We assess the effect of unknown future predictors using \[
x_t=0.7x_{t-1}+u_t,\qquad
y_t=0.8y_{t-1}-0.5y_{t-2}+x_t+\varepsilon_t,
\] where \(u_t\overset{\mathrm{iid}}{\sim}N(0,\tau^2)\) and
\(\varepsilon_t\overset{\mathrm{iid}}{\sim}N(0,1)\) are mutually
independent, with \(\tau\in\{0.25,1\}\). Each of 100 replications
retains 5000 observations after a 500-observation burn-in. At each
forecast origin, an ARX(2) model with an intercept is estimated by least
squares over a rolling window of 500 observations. We compare forecasts
using observed future predictors (Known) with forecasts using
\(\hat x_{t+j|t}=0.7^j x_t\) (Forecast). The autoregressive coefficient
of \(x_t\) is treated as known to isolate future predictor uncertainty
from its estimation error.

Both settings use the same simulated paths and the AcMCP settings in
Section~\ref{sec-simar}, but are calibrated separately. We evaluate
horizons \(h=1,2,3\) at a target coverage of \(90\%\). Local coverage
MAE measures the average absolute deviation of 500-observation rolling
coverage from \(90\%\).

Table~\ref{tbl-predictor} shows that both settings maintain coverage
close to \(90\%\), with local coverage MAE below \(0.3\) percentage
points. Interval widths in the Known setting are nearly unchanged across
the two values of \(\tau\), whereas forecasting future predictors leads
to wider intervals, with larger increases at higher \(\tau\) and longer
horizons. At \(h=3\), the increase is about \(8\%\) for \(\tau=0.25\)
and \(97\%\) for \(\tau=1\). Thus, in this simulation, AcMCP
accommodates predictor uncertainty primarily through wider intervals
while maintaining empirical coverage.

\begin{table}

\caption{\label{tbl-predictor}AcMCP with observed (Known) and forecast (Forecast) future predictors. Entries are means over 100 paired simulation replications, with Monte Carlo standard errors in parentheses. pp denotes percentage points.}

\centering{

\resizebox{\linewidth}{!}{
\begin{tabular}{lrrrrrr}
\toprule
 & \multicolumn{2}{c}{Coverage (\%)} & \multicolumn{2}{c}{Mean interval width} & \multicolumn{2}{c}{Local coverage MAE (pp)} \\
\cmidrule(l{3pt}r{3pt}){2-3} \cmidrule(l{3pt}r{3pt}){4-5} \cmidrule(l{3pt}r{3pt}){6-7}
 & \multicolumn{1}{c}{Known} & \multicolumn{1}{c}{Forecast} & \multicolumn{1}{c}{Known} & \multicolumn{1}{c}{Forecast} & \multicolumn{1}{c}{Known} & \multicolumn{1}{c}{Forecast} \\
\midrule
 & \multicolumn{6}{c}{$\tau=0.25$} \\[0.3em]
$h=1$ & 89.99 (0.003) & 89.99 (0.004) & 3.55 (0.005) & 3.65 (0.005) & 0.20 (0.001) & 0.19 (0.001) \\
$h=2$ & 89.99 (0.004) & 89.99 (0.004) & 4.72 (0.008) & 5.00 (0.010) & 0.24 (0.002) & 0.24 (0.002) \\
$h=3$ & 89.99 (0.004) & 89.99 (0.005) & 4.85 (0.009) & 5.26 (0.011) & 0.27 (0.002) & 0.27 (0.002) \\
\cmidrule(l{3pt}r{3pt}){2-7}
 & \multicolumn{6}{c}{$\tau=1$} \\[0.3em]
$h=1$ & 89.99 (0.003) & 89.99 (0.003) & 3.56 (0.005) & 4.99 (0.007) & 0.20 (0.001) & 0.19 (0.001) \\
$h=2$ & 89.99 (0.004) & 89.99 (0.004) & 4.72 (0.009) & 8.10 (0.016) & 0.24 (0.002) & 0.24 (0.001) \\
$h=3$ & 89.98 (0.005) & 89.99 (0.005) & 4.85 (0.009) & 9.57 (0.021) & 0.27 (0.002) & 0.29 (0.002) \\
\bottomrule
\end{tabular}}

}

\end{table}%

\end{document}